\documentclass[a4paper,fleqn]{cas-dc}
\pdfoutput=1
\usepackage[numbers]{natbib}

\usepackage{subcaption,tikz,float,tabularx}
\usepackage{siunitx}

\usetikzlibrary{spy,decorations.fractals}

\graphicspath{{./figure-exp3/}{./figure-exp1/}{./figure-exp2/}{./figure-exp4/}{./figure-author/}}

\definecolor{spy_color_red}{RGB}{49,120,136}    
\definecolor{spy_color_orange}{RGB}{255, 255, 255}   
\definecolor{mark_color}{RGB}{255,255,255}  

\newcommand{\acc}{ACC\,}

\begin{document}

\let\WriteBookmarks\relax
\def\floatpagepagefraction{1}
\def\textpagefraction{.001}

\shortauthors{Jiabao Wang et~al.}
\shorttitle{ }
\title [mode = title]{Align-Free Multi-Plane Phase Retrieval}  

\author[1]{Jiabao Wang}[]
\credit{Writing – original draft, Investigation, Software, Methodology, Validation, Formal analysis, Data curation, Conceptualization}

\author[1]{Yang Wu}[]
\credit{Writing – review \& editing, Methodology, Validation, Formal analysis}

\author[1]{Jun Wang}[]
\cormark[1]
\ead{jwang@scu.edu.cn}
\credit{Writing – review \& editing, Supervision, Project administration, Funding acquisition}
\affiliation[1]{organization={School of Electronics and Information Engineering},
    addressline={Sichuan University}, 
    city={Chengdu},
    postcode={610065}, 
    country={China}}
\author[2]{Ni Chen}
\cormark[1]
\ead{nichen@arizona.edu}
\credit{Writing – review \& editing, Supervision, Software, Methodology, Validation, Formal analysis,Project administration, Funding acquisition}
\affiliation[2]{organization={Wyant College of Optical Sciences},
    addressline={University of Arizona}, 
    city={Tucson},
    postcode={85721}, 
    state={AZ},
    country={USA}}

\cortext[cor1]{Corresponding author}

\begin{abstract}
The multi-plane phase retrieval method provides a budget-friendly and effective way to perform phase imaging, yet it often encounters alignment challenges due to shifts along the optical axis in experiments. Traditional methods, such as employing beamsplitters instead of mechanical stage movements or adjusting focus using tunable light sources, add complexity to the setup required for multi-plane phase retrieval. Attempts to address these issues computationally face difficulties due to the variable impact of diffraction, which renders conventional homography techniques inadequate. In our research, we introduce a novel Adaptive Cascade Calibrated (ACC) strategy for multi-plane phase retrieval that overcomes misalignment issues. This technique detects feature points within the refocused sample space and calculates the transformation matrix for neighboring planes on-the-fly to digitally adjust measurements, facilitating alignment-free multi-plane phase retrieval. This approach not only avoids the need for complex and expensive optical hardware but also simplifies the imaging setup, reducing overall costs. The effectiveness of our method is validated through simulations and real-world optical experiments.
\end{abstract}


\maketitle

\section{Introduction}\label{sec:Introduction}
Phase imagigng, the process of reconstructing the phase of a wavefront from intensity measurements, has long been a challenging task due to the ill-posed nature of the inverse problem~\cite{zuo2015transport,zuo2018phase,wang2024use,katkovnik2017computational}. 
There are numerous methods available for phase imaging, including wavefront sensors~\cite{soldevila2018phase, brady2009optical}, interferometric holography~\cite{iaconis1998spectral}, iterative phase retrieval~\cite{Zhang:03,zhang2017adaptive,Guo:21,xu2023lensless,rivenson2016sparsity}, optimization algorithms~\cite{kim2016solving}, and more. 
Among these methods, multi-plane phase retrieval stands out for its simplicity and cost-effectiveness in experimental setups, along with its straightforward iterative reconstruction algorithm. 
The experimental setup typically employs classical inline holography~\cite{goodman2005introduction,xu2001digital,garcia2006digital}, where either the sensor or the target sample translates along the optical axis to obtain various defocus measurements. Reconstruction involves iterating propagation between the sample plane and the multiple measurements, replacing the amplitude of the numerically calculated image at each measurement plane with the measured images.
However, the reconstruction process often assumes perfect alignment between the experimental setup and numerical wave propagation, requiring precise translations. Yet, translation errors are inevitable, thus limiting the quality of reconstruction in multi-plane phase retrieval.


To address the challenge of achieving precise phase imaging, the use of multiple beam splitters for inducing defocus measurements has been explored. While this approach can be effective, it often leads to a cumbersome setup that proves impractical for certain imaging configurations, such as those required for lensless imaging, as discussed in works like \cite{gao2014phase,zheng2014digital,zheng2015autofocusing,zhuo2023quantitative}. The bulkiness inherent in these configurations is a significant drawback, particularly in applications where space efficiency and simplicity are crucial.
An alternative strategy involves employing multiple measurements under varied wavelengths of light, which could potentially counter the issues associated with unstable translations intrinsic to the imaging process. This method, however, introduces the need for expensive tunable light sources, as highlighted in studies such as \cite{wu2021wavelength,luo2016pixel,luo2015synthetic,guo2021high,min2014phase}. While offering a workaround to physical limitations, the financial and operational costs associated with acquiring and integrating tunable light sources pose significant challenges.
Computational techniques present another avenue for addressing misalignment issues, offering a contrast to the hardware-intensive methods previously mentioned. However, these approaches encounter unique obstacles, particularly due to the variable effects of diffraction on imaging accuracy. The direct application of conventional computational methods like homography is hampered by these diffraction effects, making them less effective in this context, as demonstrated in research such as \cite{greenbaum2012maskless,greenbaum2014wide}.
One specific complication with computational strategies is the reliance on the use of control points or markers for feature detection. While this method facilitates certain aspects of phase retrieval, it introduces additional complexity into the experimental setup and increases the risk of sample contamination with dust. This risk is not merely theoretical; it poses a tangible threat to the integrity of the imaging process, potentially leading to compromised phase reconstruction outcomes. Furthermore, despite these considerable efforts, diffractive effects may still lead to unsuccessful phase reconstruction, underscoring the persistent challenge of managing optical phenomena in imaging applications \cite{greenbaum2012maskless,greenbaum2014wide}.
In summary, the search for efficient and effective phase retrieval techniques is characterized by a series of trade-offs between hardware complexity, operational costs, and the inherent difficulties posed by optical phenomena. The literature, including studies cited herein, reflects the ongoing exploration and development of solutions that strive to mitigate these challenges, highlighting the evolving nature of this field.

This paper introduces an Adaptive Cascade Calibrated (\acc) multi-plane phase retrieval technique. Unlike existing methods that rely on precise alignment in the experimental setup, we implement a computational self-calibration during the phase reconstruction process. Instead of directly manipulating the measurements, the \acc method identifies feature points and computes neighboring affine matrices within the refocused plane of multiple measurements. 
Subsequently, the measurements are aligned based on these calculated affine matrices before undergoing phase retrieval. Compared to existing techniques, \acc eliminates the need for markers or meticulous alignment in experimental setups, thereby preserving the simplicity and cost-effectiveness of multi-plane phase retrieval. Moreover, the calculation of affine matrices is performed in the object space rather than the measurement space, significantly reducing diffraction effects during feature detection. 
Additionally, all calculations can be automated without manual intervention. These advancements collectively enable align-free multi-plane phase retrieval.
Various experimental findings demonstrate the effectiveness of \acc  in calibrating experimental errors and achieving high-quality reconstructions. 

\section{Problem analysis}\label{sec:problem_analysis}
\begin{figure}[t]
	\centering
         \begin{tikzpicture}
		\node[anchor=south west, inner sep=0] (image) at (0,0) 
            {\includegraphics[width=1\columnwidth]{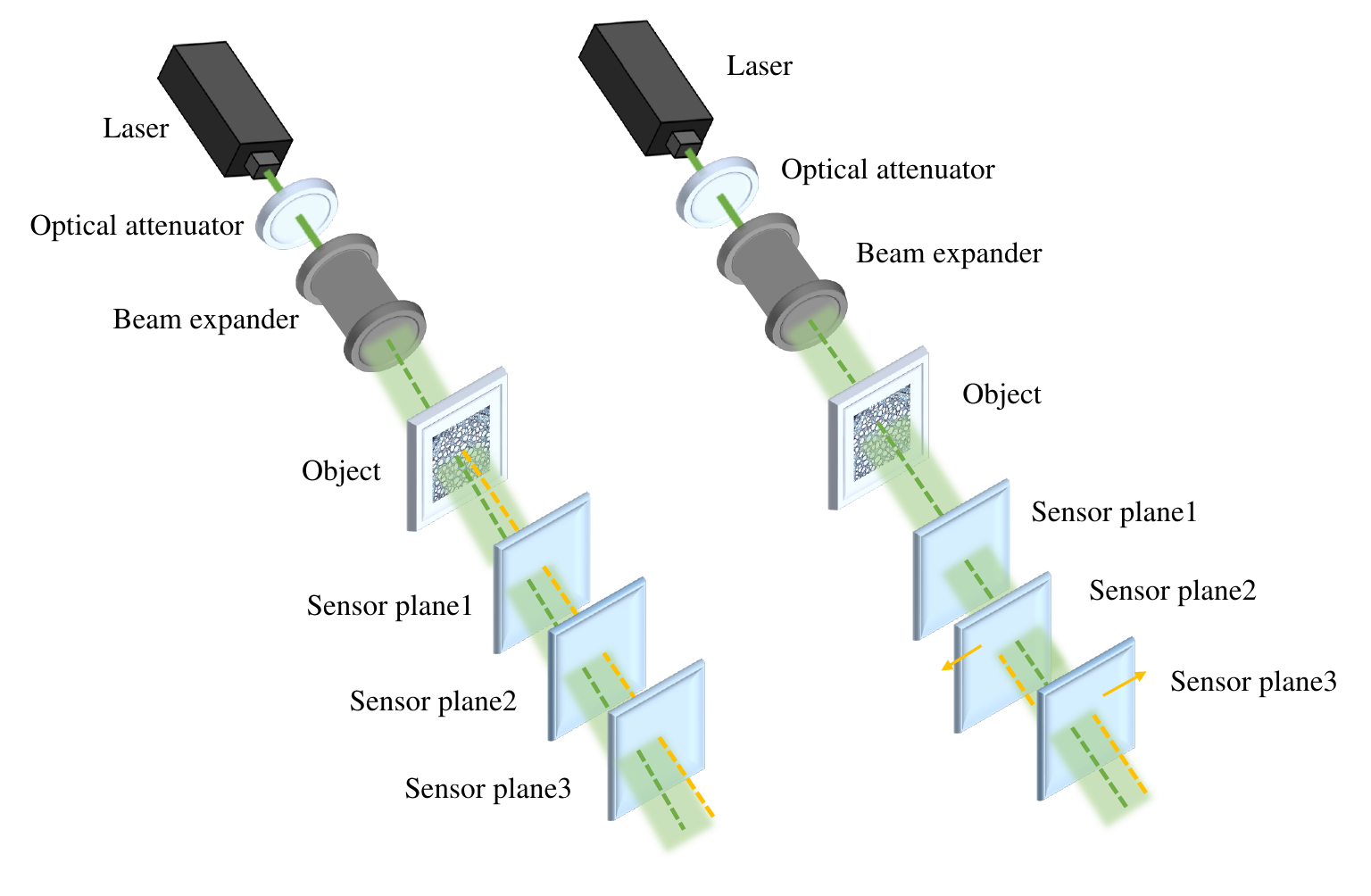}};
		\node[align=left] at (3.3, -0.1) {(a)};  
            \node[align=left] at (6.2, -0.1) {(b)};  
	\end{tikzpicture}
	\caption{Two prevalent causes of experimental errors. The left panel illustrates error cause (a), while the right panel depicts error cause (b).}
	\label{error}
\end{figure}

In prior multi-plane phase retrieval methods, despite the utilization of accurate mechanical displacement mechanisms and optical setups, errors like rotation and lateral shift among various measurement planes are inevitable due to the sensor plane's axial motion. As a result, multiple measurements must be precisely aligned for superior quality reconstruction prior to initiating the phase retrieval process.
Two typical forms of experimental error are depicted in Fig.~\ref{error}. Figure~\ref{error}(a) shows the scenario where, although the axial movement of the sensor plane is precisely managed, there's a misalignment causing the optical axis to deviate from the sensor's axis, leading to a lateral shift in the measurement plane.
Fig.~\ref{error}(b) depicts a scenario where, despite the optical axis being fixed, the sensor plane fails to move strictly along the sensor's axis, leading to lateral translation errors in the measurements that vary in direction. 

To align these measurements, traditional computational calibration approaches have often relied on manually setting feature control points \cite{greenbaum2012maskless,greenbaum2014wide}. Typically, three feature control points are chosen in one measurement (considered the reference measurement), located at different corners. Corresponding points are then identified in the other measurements. These points are aligned with the reference measurement using affine transformations.
For more accurate calculation of the feature control points' positional shifts in the measurements, these points are generally selected for their circular symmetry, such as small, isolated dust particles in the corners, or special alignment markers placed within the field of view's corners. 
The first measurement, which has the smallest distance between the sample and sensor, is often chosen as the reference plane. Subsequently, all other measurements are aligned directly to this reference plane.

\begin{figure}[t]
	\centering
        \begin{tikzpicture}
		\node[anchor=south west, inner sep=0] (image) at (0,0) {\includegraphics[width=0.95\columnwidth]{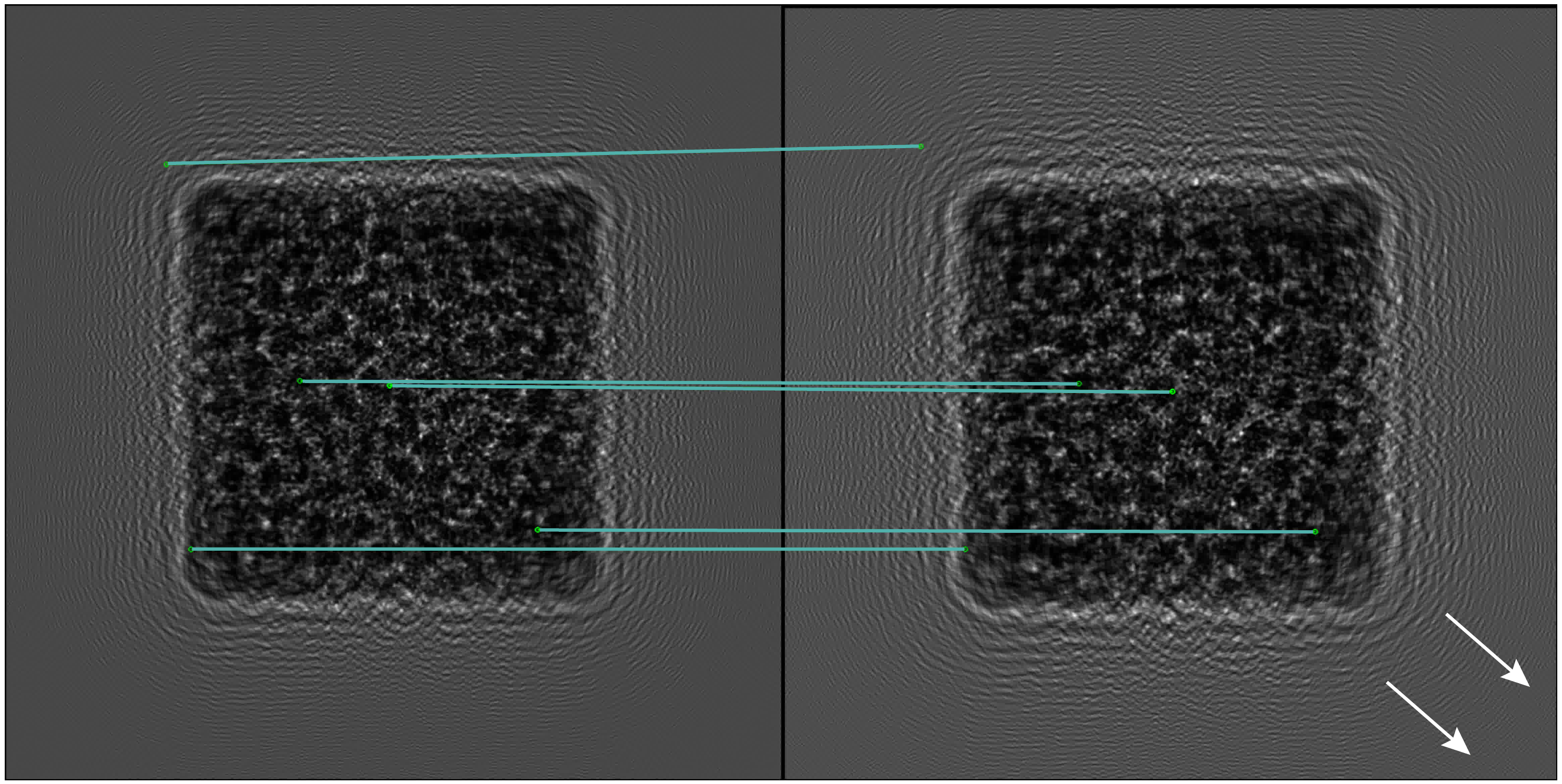}};
		\node[align=left, white] at (2, 3.75) {Measurement 1};  
            \node[align=left, white] at (5.9, 3.75) {Measurement 2};  
	\end{tikzpicture}
	\caption{Feature point detection between the measurements.}
	\label{Detecting_holo}
\end{figure}

However, the reliability of manually chosen feature control points, positioned at the sample's edges, diminishes as the magnitude of lateral translation error increases. These points may drift beyond the sensor's field of view, rendering them ineffective for calibration purposes. 
Furthermore, as the separation between planes grows, the resulting diffraction effect within the measurements becomes more pronounced, complicating the task of identifying distinct control points across different measurements. 
The challenge of detecting feature control points within measurements using previous multi-plane calibration techniques is illustrated in Fig.~\ref{Detecting_holo}, as indicated by arrows showing a shift towards the lower right in plane 2. The cyan lines depict the feature points and their detected mappings using the conventional approach, which directly applies homography to the diffracted measurements. Ideally, these lines connecting feature points should be parallel. However, the depicted matching lines in cyan fail to demonstrate this parallelism, underscoring the difficulties in distinguishing feature points between two measurements due to enhanced diffraction effects. Moreover, the act of manually placing markers on the sample could potentially damage it and complicate the experimental setup.

\section{Method}\label{sec:Method}
\begin{figure*}[!t]
	\centering
	\includegraphics[width=2\columnwidth]{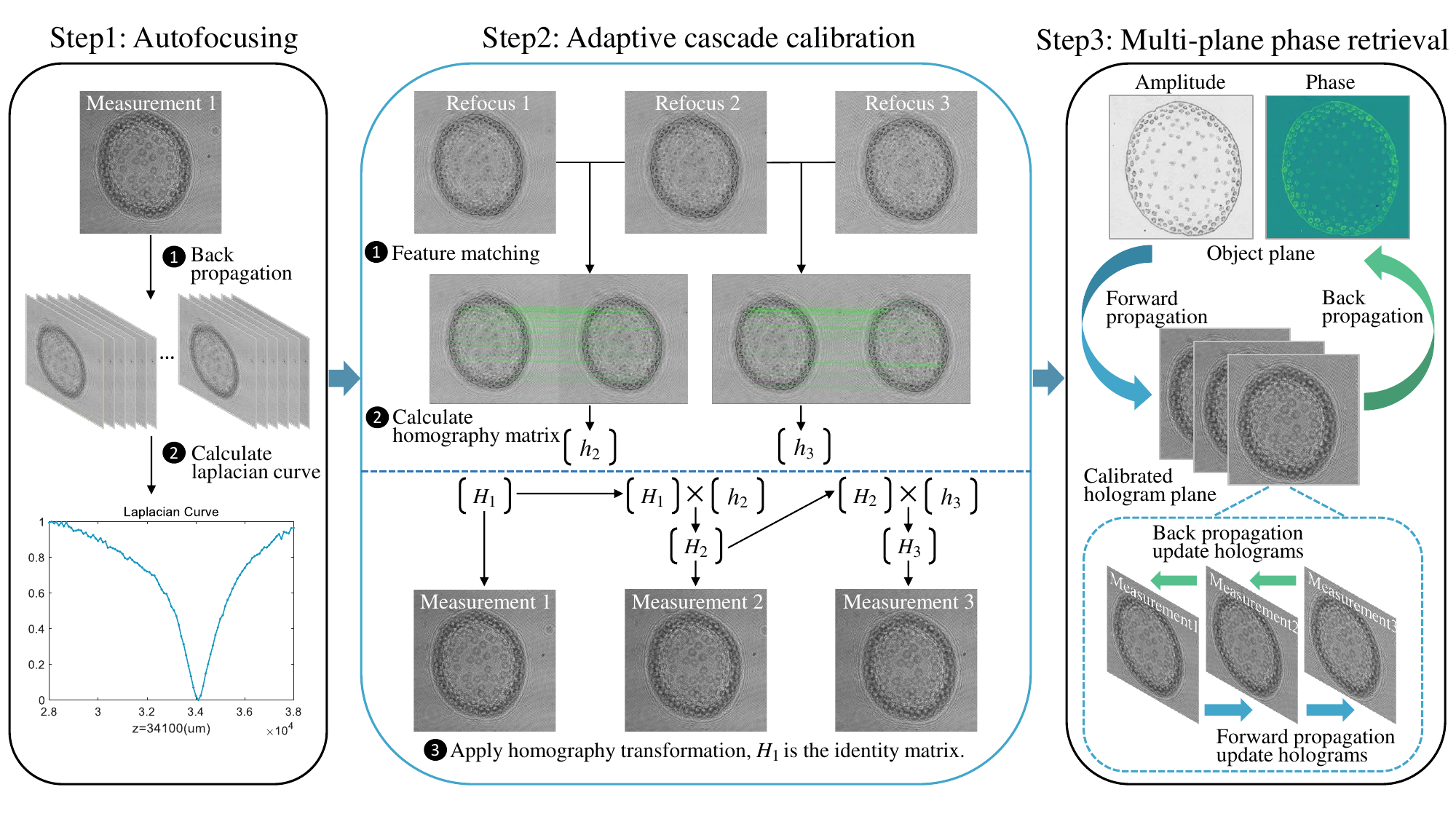}
	\caption{Schematic diagram of multi-plane phase retrieval algorithm based on adaptive cascade calibrated algorithm.}
	\label{calibration}
\end{figure*}

To address the challenges highlighted earlier, our approach introduces a novel, experimentally align-free multi-plane phase retrieval strategy that encompasses three key components: autofocusing, adaptive cascade calibration, and multi-plane phase retrieval itself. This method is designed to streamline the phase retrieval process, eliminating the need for complex alignment procedures and thereby simplifying experimental setups. Here, we expand upon each of these integral elements, providing a comprehensive overview of their functionalities and contributions to the overall method.

\subsection{Autofocusing algorithm}
Autofocusing is performed for two primary purposes: 1) to accurately determine the locations of each set of measurements; 2) to acquire a refocused image of the target sample for the purpose of calibration.
We adopt the Laplace operator as the criterion, where the minimum value indicates a sharp reconstruction and precise measurement distance~\cite{Langehanenberg2008,PechPacheco2000,Mudanyali2010}. The Laplace operator of an image $I$ is provided by
\begin{equation}
\label{Laplace}
\Delta I(m,n) = \sum_{m}^{M}\sum_{n}^{N} |\nabla^2 I(m,n)|.
\end{equation}
Here, $\nabla^2 I(m,n)$ represents the second-order derivative of the input image $I(m,n)$, with the final outcome derived by summing the absolute values of this derivative.
Back-propagation is executed across a range of propagation distances, utilizing the Laplacian metric to gauge the degree of focus. The lowest value pinpointed by this metric signifies the precise location of the corresponding measurement. This procedure is applied to each set of measurements. The back-propagated images, identified at the optimal focusing distance, are then prepared for the subsequent stage of the process, as depicted in Step 1 of Fig. \ref{calibration}.

\subsection{Adaptive cascade calibration algorithm}
Utilizing the refocused images acquired for each measurement from the initial step, we employ the Scale Invariant Feature Transform (SIFT) algorithm~\cite{Lowe1999, Lowe2004} to identify feature points between two adjacent measurements, complemented by a fast approximate nearest neighbor matcher for pairing these feature points. To ensure the highest quality of matches, Lowe's algorithm~\cite{Lowe2004} aids in filtering out outlier data points within the matches. Building on this, we employ a matcher to compute the homography matrix $h_n$ between two adjacent object planes, as illustrated in Step 2.
The affine matrix $H_n$ between the first and the $n$-th measurement is calculated as $H_n = H_{n-1} \times h_n$. It is evident that $H_1=h_1$, both representing the identity matrix. This technique of employing an indirect, cascading approach for the affine matrix computation is designed to overcome the difficulties encountered in detecting significant transformations.
The computed affine matrices are subsequently utilized to digitally calibrate each measurement.

\subsection{Multi-plane phase retrieval algorithm}
We conduct a energy-conserved Gerchberg-Saxton (GS) \cite{GS} algorithmic to perform the final phase retrieval, which uses numerical diffraction calculation to propagate back and forth in different measurement planes in order to recover the phase information of object from intensity-only measurements.
As depicted in Step 3 of Fig.~\ref{calibration}, The process begins by setting the initial phase distribution on the object plane to zero. The complex field of the initial object is forward propagated to each measurement location, where their amplitude are 
replaced by the square roots of the calibrated measurements retrievaled in step 2 and the phase are maintained. This reverse propagation process at the object plane is subject to an energy-conserving constraint \cite{Latychevskaia2007}, acknowledging that the sample object does not exhibit negative absorption. This constraint significantly reduces the number of required measurements and accelerates the convergence of the iterative process, allowing high-quality phase reconstruction with just three measurements.
\section{Results}\label{sec:Experiment}

\subsection{Numerical Experiments}
\begin{figure}[h]
	\centering
         \begin{tikzpicture}
		\node[anchor=south west, inner sep=0] (image) at (0,0) {\includegraphics[width=1\columnwidth]{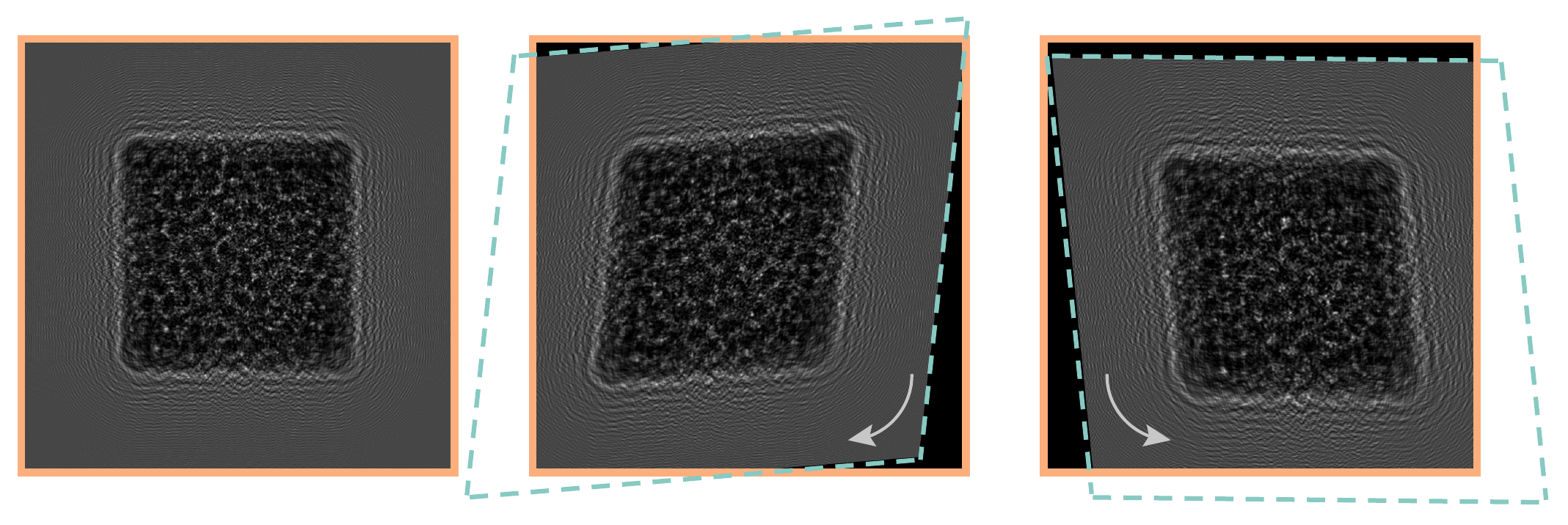}};
		\node[align=left] at (1.2, 2.9) {Plane 1};  
            \node[align=left] at (3.9, 2.9) {Plane 2};  
            \node[align=left] at (6.7, 2.9) {Plane 3};  
	\end{tikzpicture}
	\caption{Measurements with and without simulated misalignment.}
	\label{holo_error}
\end{figure}
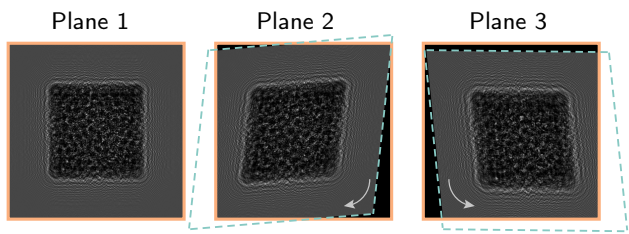

To evaluate the effectiveness of our proposed method, we begin with simulations. 
We simulate the multiple plane measurements with an inline setup as shown in Fig.~\ref{error}. The illumination light source was a laser with wavelength centered at \SI{532}{\nano\meter}, and the camera sensor was with pixel pitch of \SI{3.45}{\micro\meter}. 
The band-limited angular spectrum method was utilized to compute measurements at various planes, and Gaussian noise with a signal-to-noise ratio of 30 dB was introduced to each measurement. 
We selected three measurements at sample-to-sensor distances of \SI{30}{\milli\meter}, \SI{40}{\milli\meter}, and \SI{50}{\milli\meter}.
To mimic misalignment in actual experiments, we create $3 \times 3$ randomized transformation matrices and apply these transformations to each measurement using Eq.~\ref{matrix}. 
The vectors $[u, v, 1]^T$ and $[x, y, z]^T$ denote the coordinates before and after transformation, respectively. Within these matrices, the parameters $a_{13}$ and $a_{23}$ dictate translational errors, $a_{12}$ and $a_{21}$ manage rotational errors, and $a_{11}$ and $a_{22}$ adjust for scaling errors.
Fig.~\ref{holo_error} presents an illustration of measurements without misalignment, alongside measurements that exhibit misalignments induced by the described random transformation matrices.
\begin{equation}
	\label{matrix}
         \begin{bmatrix} 
		x \\ 
		y \\
		z
	\end{bmatrix}
        =
	\begin{bmatrix} 
		a_{11} & a_{12} & a_{13} \\ 
		a_{21} & a_{22} & a_{23} \\
		a_{31} & a_{32} & a_{33}
	\end{bmatrix}
        \times
        \begin{bmatrix} 
		u \\ 
		v \\
		1
	\end{bmatrix}
\end{equation}

\begin{figure*}[h]
	\centering
	\begin{subfigure}[t]{2\columnwidth}
		\centering  
		\begin{tikzpicture}[spy using outlines={rectangle, width=0.9cm, height=0.9cm, spy_color_red, magnification=3, connect spies}]
			\node[anchor=south west, inner sep=0] (image) at (0,0) {\includegraphics[width=0.21\columnwidth]{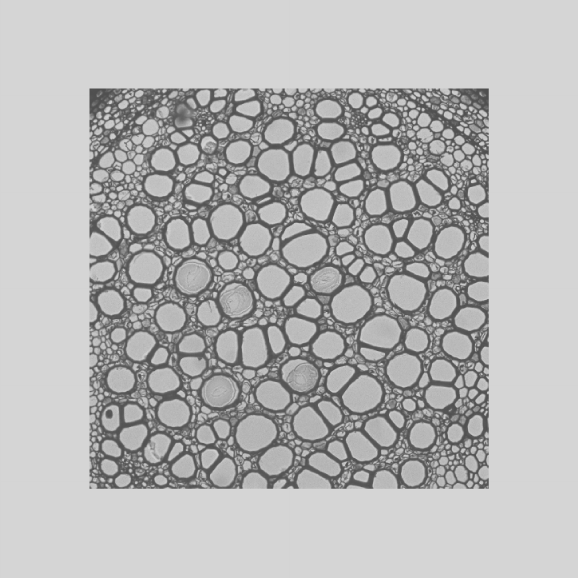}};
			\spy[spy_color_red, opacity=1, line width=0.5pt] on (2.15, 1.5) in node[anchor=south east, line width=0.5pt] at (3.5, 0.04);
			\spy[spy_color_red, opacity=1, line width=0.5pt] on (1.05, 1.6) in node[anchor=north west, line width=0.5pt] at (0.04, 3.5);
			\node[align=left] at (1.6, 3.75) {amplitude};  
		\end{tikzpicture}
		\hspace{0.001\textwidth}       
		\begin{tikzpicture}[spy using outlines={rectangle, width=0.9cm, height=0.9cm, spy_color_red, magnification=3, connect spies}] 
			\node[anchor=south west, inner sep=0] (image) at (0,0) {\includegraphics[width=0.21\columnwidth]{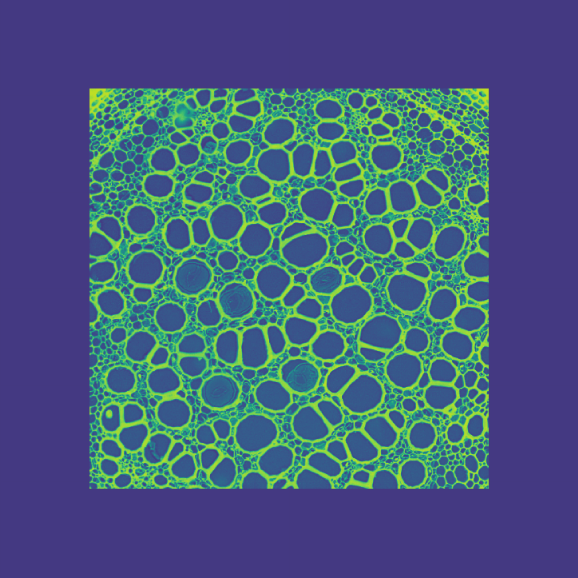}};
			\spy[spy_color_orange, opacity=1, line width=0.5pt] on (2.15, 1.5) in node[anchor=south east, line width=0.5pt] at (3.5, 0.04);
			\spy[spy_color_orange, opacity=1, line width=0.5pt] on (1.05, 1.6) in node[anchor=north west, line width=0.5pt] at (0.04, 3.5);
			\node[align=left] at (1.6, 3.75) {phase};  
		\end{tikzpicture} 
		\hspace{0.001\textwidth} 
		\begin{tikzpicture}[spy using outlines={rectangle, width=0.9cm, height=0.9cm, spy_color_red, magnification=3, connect spies}]
			\node[anchor=south west, inner sep=0] (image) at (0,0) {\includegraphics[width=0.21\columnwidth]{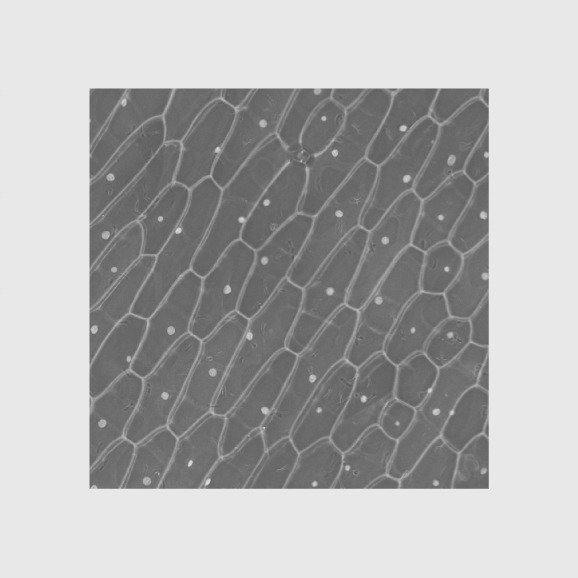}};
			\spy[spy_color_red, opacity=1, line width=0.5pt] on (2.35, 1.7) in node[anchor=south east, line width=0.5pt] at (3.5, 0.04);
			\spy[spy_color_red, opacity=1, line width=0.5pt] on (1.06, 2.14) in node[anchor=north west, line width=0.5pt] at (0.04, 3.5);
			\node[align=left] at (1.6, 3.75) {amplitude};  
		\end{tikzpicture}
		\hspace{0.001\textwidth}       
		\begin{tikzpicture}[spy using outlines={rectangle, width=0.9cm, height=0.9cm, spy_color_red, magnification=3, connect spies}] 
			\node[anchor=south west, inner sep=0] (image) at (0,0) {\includegraphics[width=0.21\columnwidth]{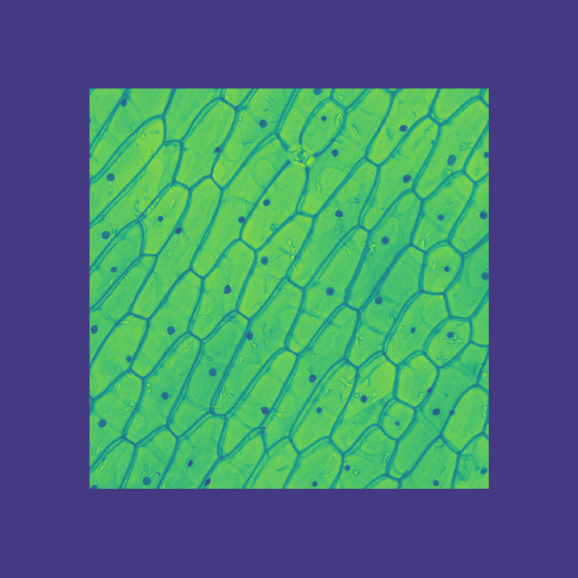}};
			\spy[spy_color_orange, opacity=1, line width=0.5pt] on (2.35, 1.7) in node[anchor=south east, line width=0.5pt] at (3.5, 0.04);
			\spy[spy_color_orange, opacity=1, line width=0.5pt] on (1.06, 2.14) in node[anchor=north west, line width=0.5pt] at (0.04, 3.5);
			\node[align=left] at (1.6, 3.75) {phase};  
		\end{tikzpicture}
		\includegraphics[height=0.21\columnwidth]{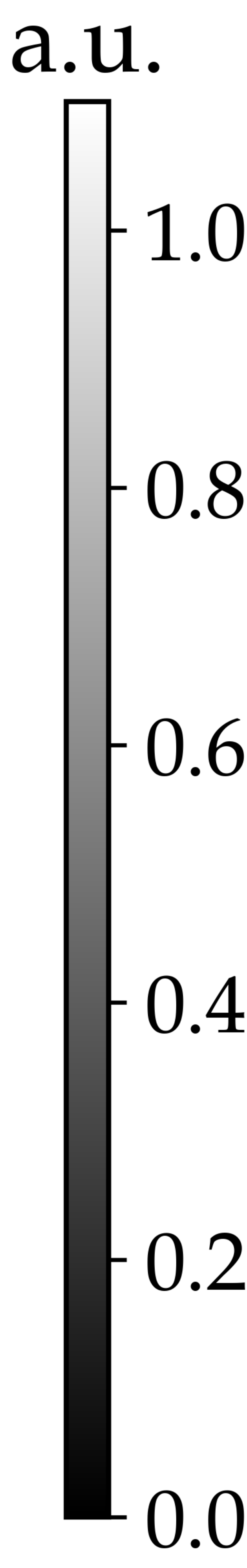} 
		\includegraphics[height=0.21\columnwidth]{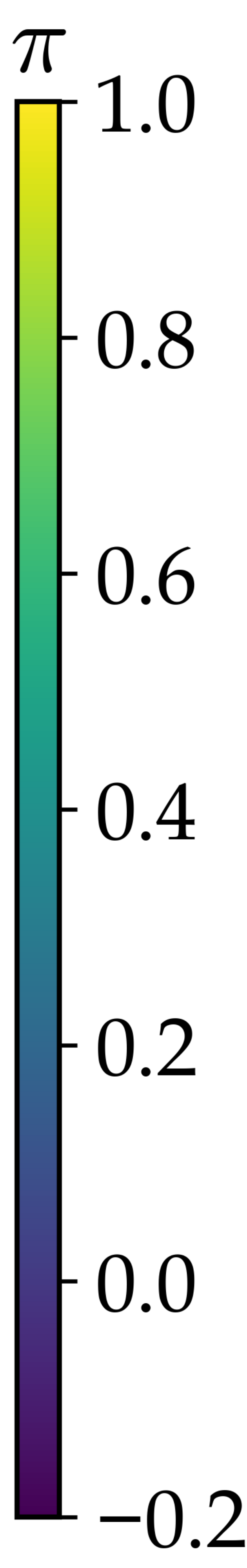}     
		\caption{Amplitude and phase of the target samples (cell crystals and onion epidermal cells)}
		\label{subfig:Experiment1a}
	\end{subfigure}

	\begin{subfigure}[t]{2\columnwidth}
		\centering  
		\begin{tikzpicture}[spy using outlines={rectangle, width=0.9cm, height=0.9cm, spy_color_red, magnification=3, connect spies}]
			\node[anchor=south west, inner sep=0] (image) at (0,0) {\includegraphics[width=0.21\columnwidth]{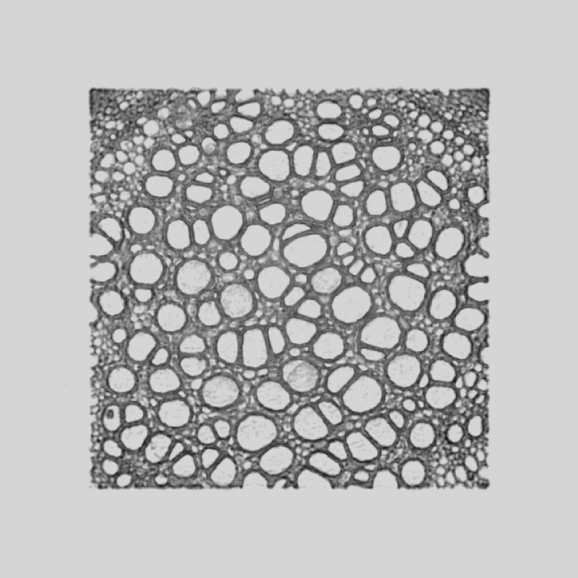}};
			\spy[spy_color_red, opacity=1, line width=0.5pt] on (2.15, 1.5) in node[anchor=south east, line width=0.5pt] at (3.5, 0.04);
			\spy[spy_color_red, opacity=1, line width=0.5pt] on (1.05, 1.6) in node[anchor=north west, line width=0.5pt] at (0.04, 3.5);
		\end{tikzpicture}
		\hspace{0.001\textwidth}          
		\begin{tikzpicture}[spy using outlines={rectangle, width=0.9cm, height=0.9cm, spy_color_red, magnification=3, connect spies}] 
			\node[anchor=south west, inner sep=0] (image) at (0,0) {\includegraphics[width=0.21\columnwidth]{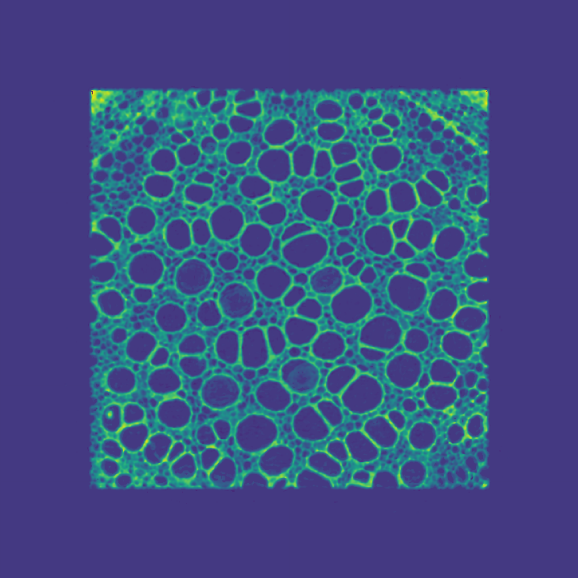}};
			\spy[spy_color_orange, opacity=1, line width=0.5pt] on (2.15, 1.5) in node[anchor=south east, line width=0.5pt] at (3.5, 0.04);
			\spy[spy_color_orange, opacity=1, line width=0.5pt] on (1.05, 1.6) in node[anchor=north west, line width=0.5pt] at (0.04, 3.5);
		\end{tikzpicture}
		\hspace{0.001\textwidth}      
		\begin{tikzpicture}[spy using outlines={rectangle, width=0.9cm, height=0.9cm, spy_color_red, magnification=3, connect spies}]
			\node[anchor=south west, inner sep=0] (image) at (0,0) {\includegraphics[width=0.21\columnwidth]{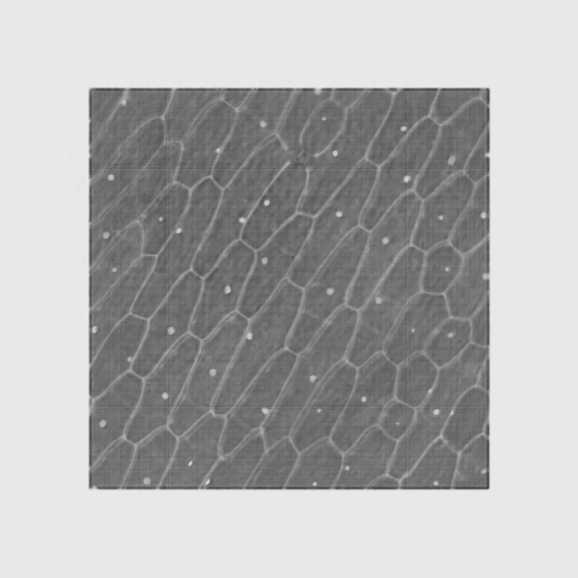}};
			\spy[spy_color_red, opacity=1, line width=0.5pt] on (2.35, 1.7) in node[anchor=south east, line width=0.5pt] at (3.5, 0.04);
			\spy[spy_color_red, opacity=1, line width=0.5pt] on (1.06, 2.14) in node[anchor=north west, line width=0.5pt] at (0.04, 3.5);
		\end{tikzpicture}
		\hspace{0.001\textwidth}          
		\begin{tikzpicture}[spy using outlines={rectangle, width=0.9cm, height=0.9cm, spy_color_red, magnification=3, connect spies}] 
			\node[anchor=south west, inner sep=0] (image) at (0,0) {\includegraphics[width=0.21\columnwidth]{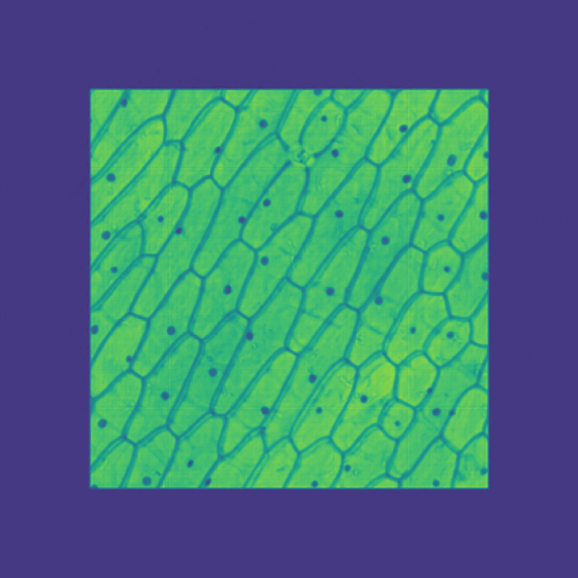}};
			\spy[spy_color_orange, opacity=1, line width=0.5pt] on (2.35, 1.7) in node[anchor=south east, line width=0.5pt] at (3.5, 0.04);
			\spy[spy_color_orange, opacity=1, line width=0.5pt] on (1.06, 2.14) in node[anchor=north west, line width=0.5pt] at (0.04, 3.5); 
		\end{tikzpicture}
		\includegraphics[height=0.21\columnwidth]{colorbar-amp-exp1} 
		\includegraphics[height=0.21\columnwidth]{colorbar-phase-exp1}      
		\caption{ACC reconstructions}
		\label{subfig:Experiment1b}
	\end{subfigure}
	
	\begin{subfigure}[t]{2\columnwidth}
		\centering
		\begin{tikzpicture}[spy using outlines={rectangle, width=0.9cm, height=0.9cm, spy_color_red, magnification=3, connect spies}]
			\node[anchor=south west, inner sep=0] (image) at (0,0) {\includegraphics[width=0.21\columnwidth]{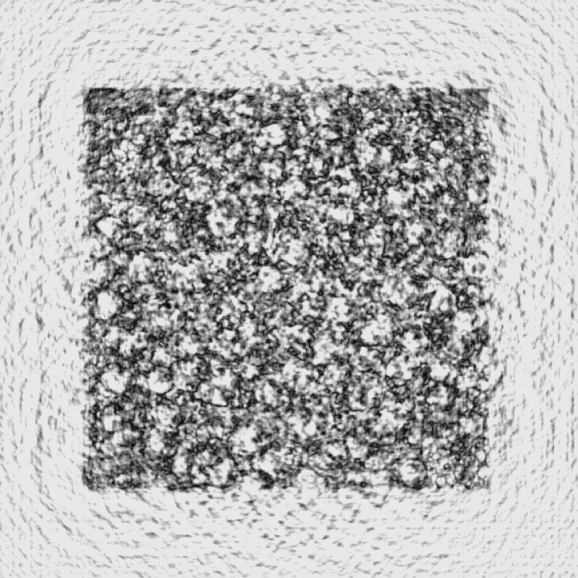}};
			\spy[spy_color_red, opacity=1, line width=0.5pt] on (2.15, 1.5) in node[anchor=south east, line width=0.5pt] at (3.5, 0.04);
			\spy[spy_color_red, opacity=1, line width=0.5pt] on (1.05, 1.6) in node[anchor=north west, line width=0.5pt] at (0.04, 3.5); 
		\end{tikzpicture}
		\hspace{0.001\textwidth}     
		\begin{tikzpicture}[spy using outlines={rectangle, width=0.9cm, height=0.9cm, spy_color_red, magnification=3, connect spies}]
			\node[anchor=south west, inner sep=0] (image) at (0,0) {\includegraphics[width=0.21\columnwidth]{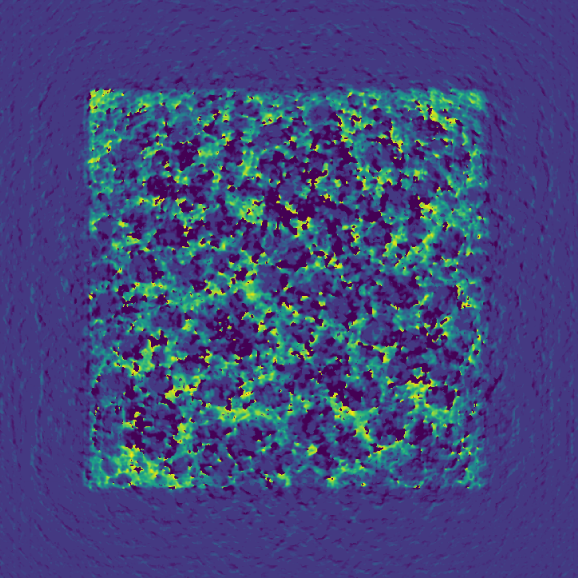}};
			\spy[spy_color_orange, opacity=1, line width=0.5pt] on (2.15, 1.5) in node[anchor=south east, line width=0.5pt] at (3.5, 0.04);
			\spy[spy_color_orange, opacity=1, line width=0.5pt] on (1.05, 1.6) in node[anchor=north west, line width=0.5pt] at (0.04, 3.5);   
		\end{tikzpicture} 
		\hspace{0.001\textwidth}    
		\begin{tikzpicture}[spy using outlines={rectangle, width=0.9cm, height=0.9cm, spy_color_red, magnification=3, connect spies}]
			\node[anchor=south west, inner sep=0] (image) at (0,0) {\includegraphics[width=0.21\columnwidth]{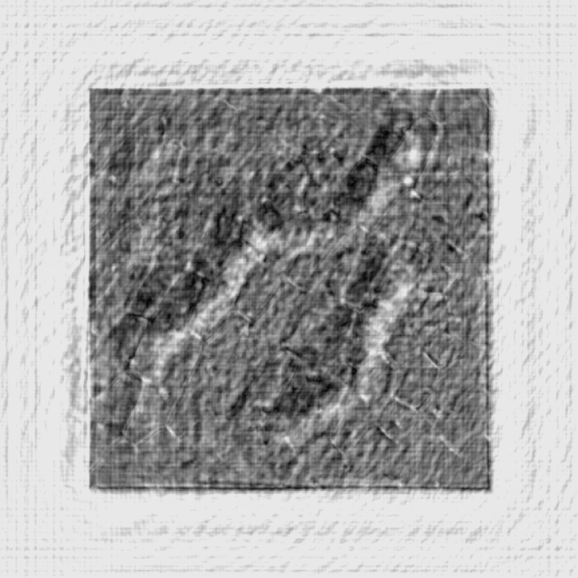}};
			\spy[spy_color_red, opacity=1, line width=0.5pt] on (2.35, 1.7) in node[anchor=south east, line width=0.5pt] at (3.5, 0.04);
			\spy[spy_color_red, opacity=1, line width=0.5pt] on (1.06, 2.14) in node[anchor=north west, line width=0.5pt] at (0.04, 3.5);
		\end{tikzpicture}
		\hspace{0.001\textwidth}     
		\begin{tikzpicture}[spy using outlines={rectangle, width=0.9cm, height=0.9cm, spy_color_red, magnification=3, connect spies}]
			\node[anchor=south west, inner sep=0] (image) at (0,0) {\includegraphics[width=0.21\columnwidth]{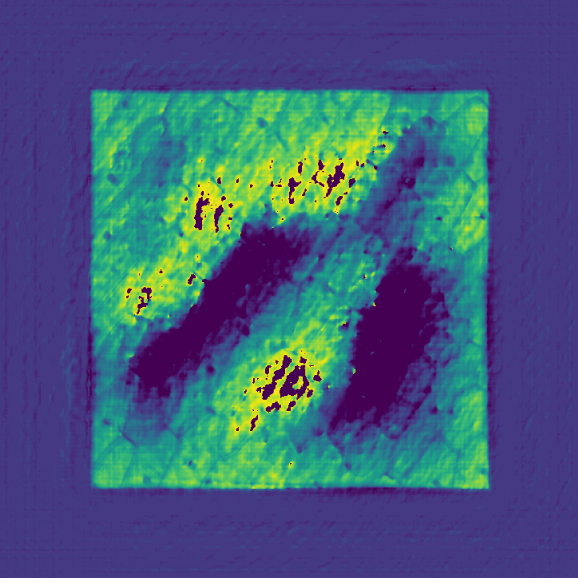}};
			\spy[spy_color_orange, opacity=1, line width=0.5pt] on (2.35, 1.7) in node[anchor=south east, line width=0.5pt] at (3.5, 0.04);
			\spy[spy_color_orange, opacity=1, line width=0.5pt] on (1.06, 2.14) in node[anchor=north west, line width=0.5pt] at (0.04, 3.5);    
		\end{tikzpicture}  
		\includegraphics[height=0.21\columnwidth]{colorbar-amp-exp1} 
		\includegraphics[height=0.21\columnwidth]{colorbar-phase-exp1}     
		\caption{Nai\"ve reconstructions}
		\label{subfig:Experiment1c}
	\end{subfigure}
	\caption{Numerical Experiment Comparison: ACC vs. Nai\"ve Methods.}
	\label{fig:Experiment1}
\end{figure*}
We conducted two group of tests on crystals and onion epidermal cells.
Figure~\ref{fig:Experiment1}(a) displays both the amplitude and phase of the simulated crystals and onion epidermal cells. We apply the ACC method outlined earlier to these measurements for reconstruction purposes. 
The reconstructions, as demonstrated in Fig. \ref{fig:Experiment1}(b), reveal that the amplitudes and phases closely align with those of the actual samples. 
In contrast, reconstructions performed by a straightforward multiple phase retrieval approach, without accounting for misalignment, are depicted in Fig.~\ref{fig:Experiment1}(c). 
These reconstructions are unsuccessful, primarily due to misalignment in the measurements. 
To thoroughly evaluate the reconstruction quality, we employ the peak signal-to-noise ratio~(PSNR) and structural similarity index~(SSIM) as metrics, comparing the reconstructed amplitude to the original amplitude. 
The results, listed in Table~\ref{table1}, indicate that the reconstructions of the proposed method exhibit significantly improved PSNR and SSIM values compared to those reconstructed without considering misalignment (Nai\"ve). 
This evidence confirms the capability of our proposed method to effectively correct random errors in measurements and achieve precise phase retrieval.

\begin{table}[h]
	\caption{PSNR, SSIM of the reconstructions in Fig.~\ref{fig:Experiment1}.}\label{table1}
	\centering
	\begin{tabular}{cccccc}
		\hline
		& \multicolumn{2}{c}{Na\"ive reconstruction} & \multicolumn{2}{c}{ACC reconstruction} \\
		\cline{2-5}
		              & PSNR/dB & SSIM & PSNR/dB & SSIM \\
		\hline
		cell crystals & 13.14 & 0.30 & 20.25 & 0.81 \\
		
		onion cells & 18.47 & 0.43 & 28.62 & 0.82 \\
		\hline
	\end{tabular}
\end{table}

\begin{figure}[h]
	\centering
 
	\begin{subfigure}[t]{1\columnwidth}
		\centering
            \includegraphics[width=0.9\columnwidth]{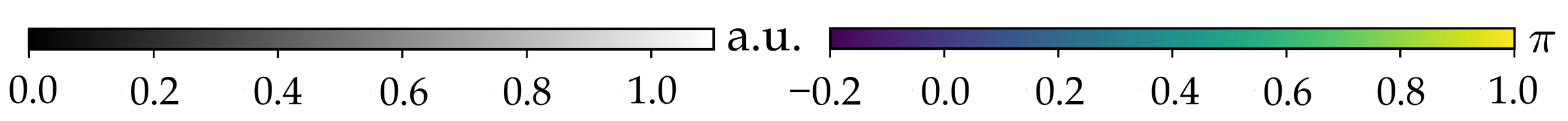}  
		\begin{tikzpicture}
			\node[anchor=south west, inner sep=0] (image) at (0,0) {\includegraphics[width=0.24\columnwidth]{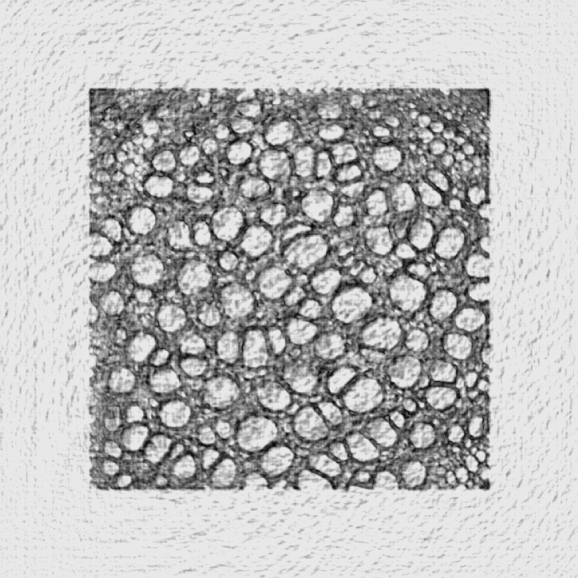}};
			\node[align=left] at (1, 2.2) {amplitude};  
		\end{tikzpicture}  
		\begin{tikzpicture}
			\node[anchor=south west, inner sep=0] (image) at (0,0) {\includegraphics[width=0.24\columnwidth]{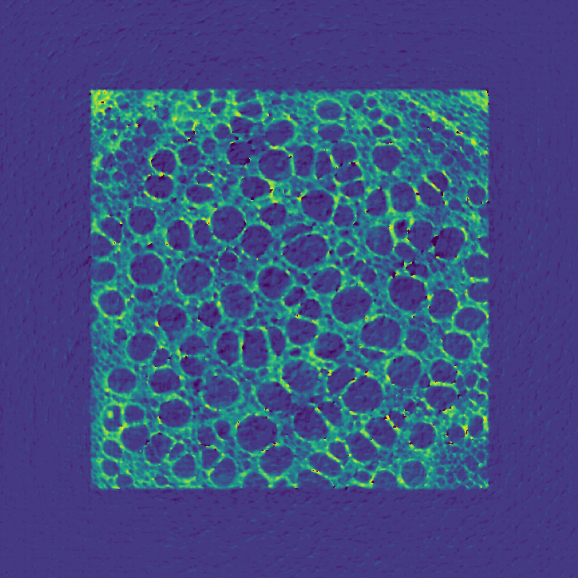}};
			\node[align=left] at  (1, 2.2) {phase};  
		\end{tikzpicture}     
		\begin{tikzpicture}
			\node[anchor=south west, inner sep=0] (image) at (0,0) {\includegraphics[width=0.24\columnwidth]{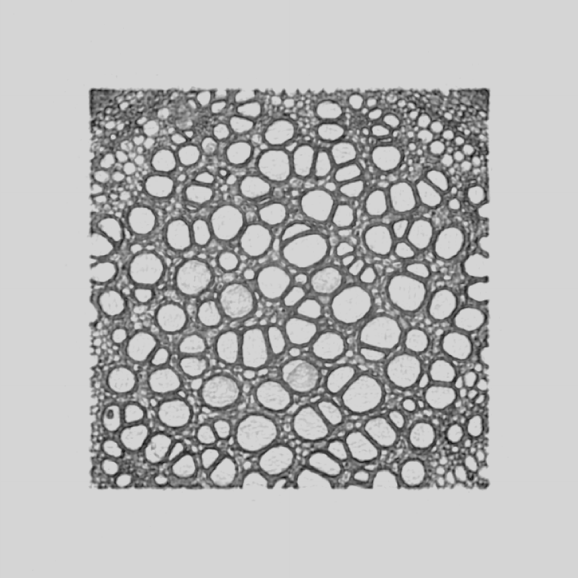}};
			\node[align=left] at  (1, 2.2) {amplitude};  
		\end{tikzpicture}     
		\begin{tikzpicture}
			\node[anchor=south west, inner sep=0] (image) at (0,0) {\includegraphics[width=0.24\columnwidth]{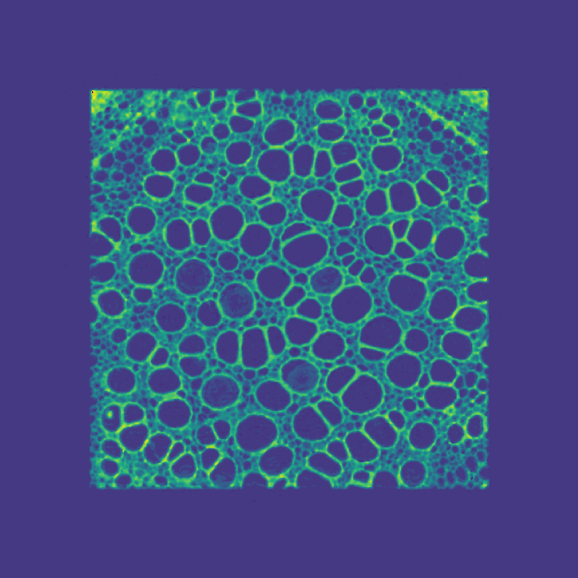}};
			\node[align=left] at  (1, 2.2) {phase};  
		\end{tikzpicture}   		
		\caption{Comparison results are presented for translation errors of 0 pixels, 2 pixels, and 4 pixels across the three planes.}
		\label{subfig:exp21}
	\end{subfigure}
 
	\begin{subfigure}[t]{1\columnwidth}
		\centering  
		\begin{tikzpicture}
			\node[anchor=south west, inner sep=0] (image) at (0,0) {\includegraphics[width=0.24\columnwidth]{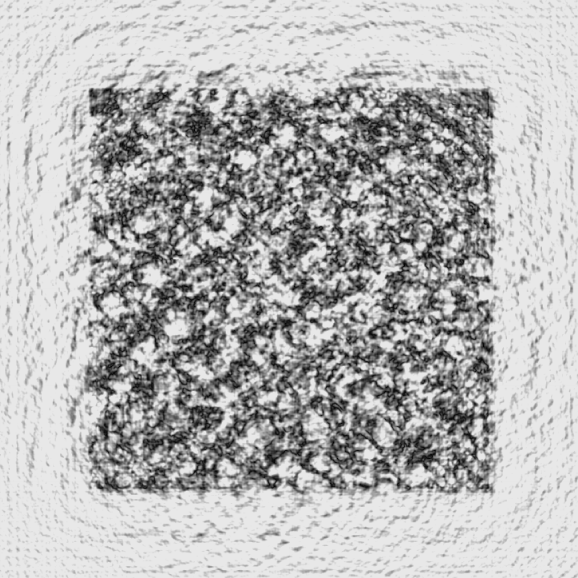}};
		\end{tikzpicture}     
		\begin{tikzpicture}
			\node[anchor=south west, inner sep=0] (image) at (0,0) {\includegraphics[width=0.24\columnwidth]{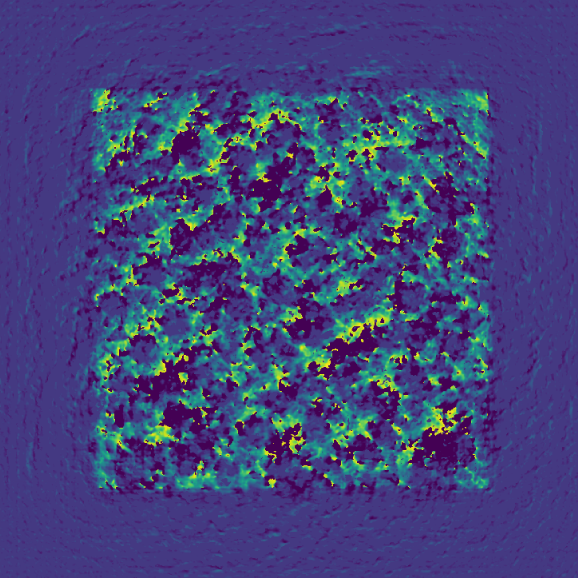}}; 
		\end{tikzpicture}      
		\begin{tikzpicture}
			\node[anchor=south west, inner sep=0] (image) at (0,0) {\includegraphics[width=0.24\columnwidth]{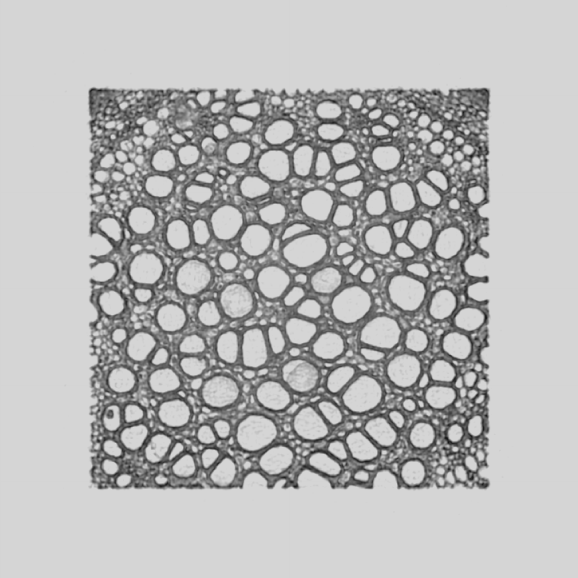}};
		\end{tikzpicture}     
		\begin{tikzpicture}
			\node[anchor=south west, inner sep=0] (image) at (0,0) {\includegraphics[width=0.24\columnwidth]{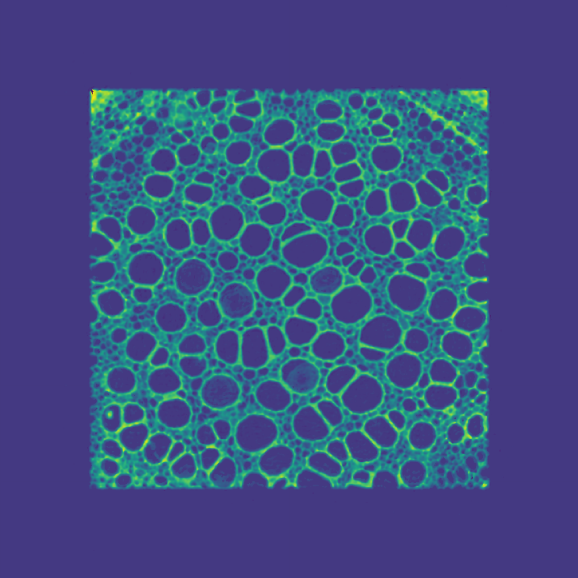}};
		\end{tikzpicture}  		
		\caption{Comparison results are presented for translation errors of 0 pixels, 5 pixels, and 10 pixels across the three planes.}
		\label{subfig:exp22}
	\end{subfigure}	
 
	\begin{subfigure}[t]{1\columnwidth}
		\centering  
		\begin{tikzpicture}
			\node[anchor=south west, inner sep=0] (image) at (0,0) {\includegraphics[width=0.24\columnwidth]{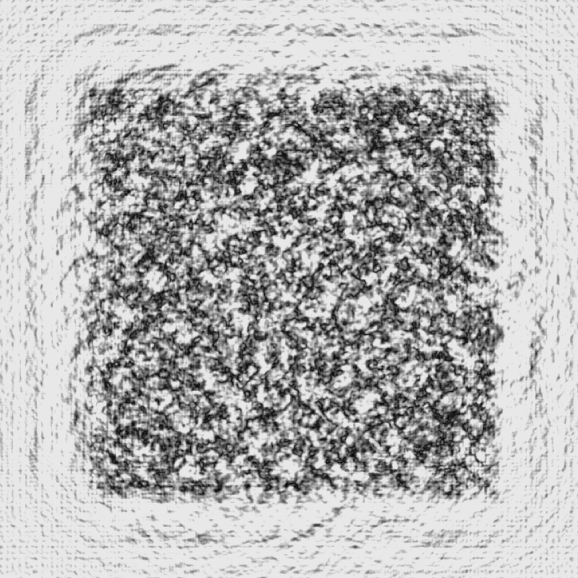}}; 
		\end{tikzpicture}      
		\begin{tikzpicture}
			\node[anchor=south west, inner sep=0] (image) at (0,0) {\includegraphics[width=0.24\columnwidth]{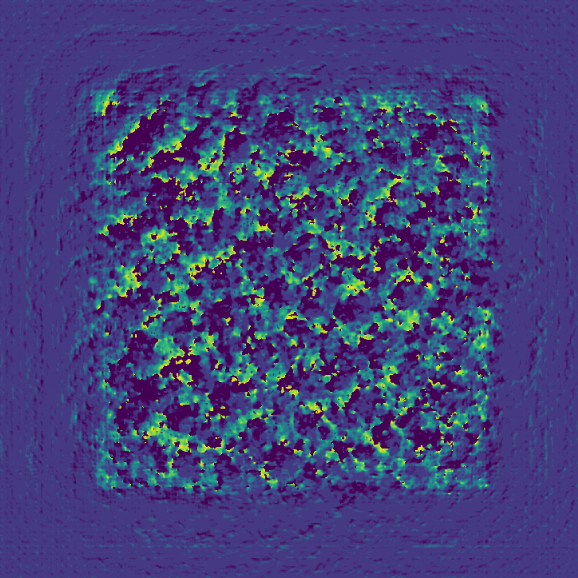}};
		\end{tikzpicture}      
		\begin{tikzpicture}
			\node[anchor=south west, inner sep=0] (image) at (0,0) {\includegraphics[width=0.24\columnwidth]{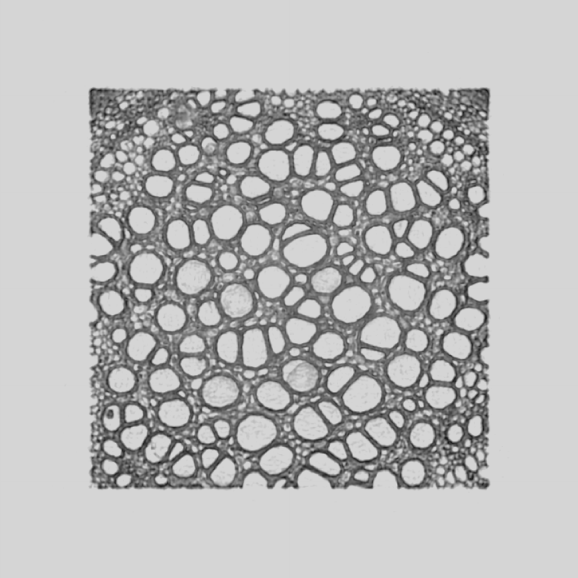}};
		\end{tikzpicture}     
		\begin{tikzpicture}
			\node[anchor=south west, inner sep=0] (image) at (0,0) {\includegraphics[width=0.24\columnwidth]{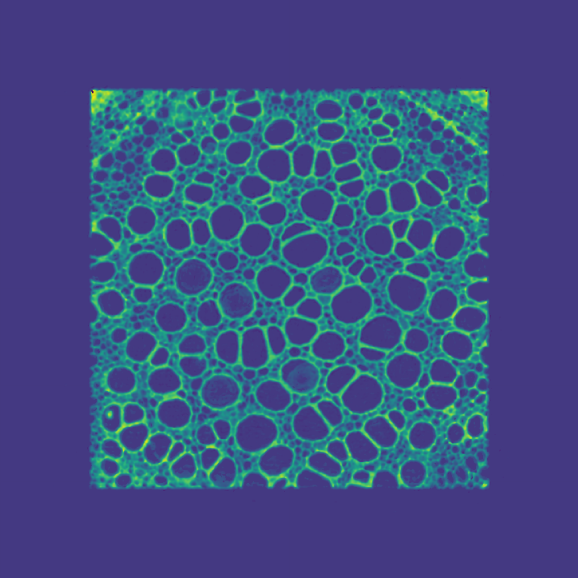}};
		\end{tikzpicture}  		
		\caption{Comparison results are presented for translation errors of 0 pixels, 8 pixels, and 16 pixels across the three planes.}
		\label{subfig:exp23}
	\end{subfigure}

	\begin{subfigure}[t]{1\columnwidth}
		\centering  
		\begin{tikzpicture}
			\node[anchor=south west, inner sep=0] (image) at (0,0) {\includegraphics[width=0.24\columnwidth]{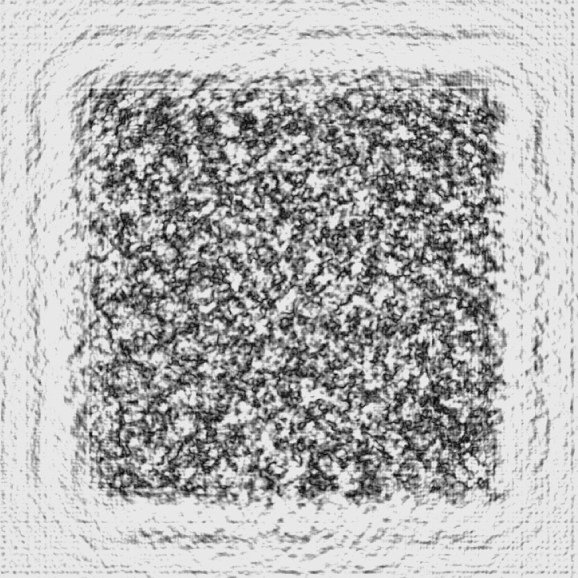}}; 
		\end{tikzpicture}     
		\begin{tikzpicture}
			\node[anchor=south west, inner sep=0] (image) at (0,0) {\includegraphics[width=0.24\columnwidth]{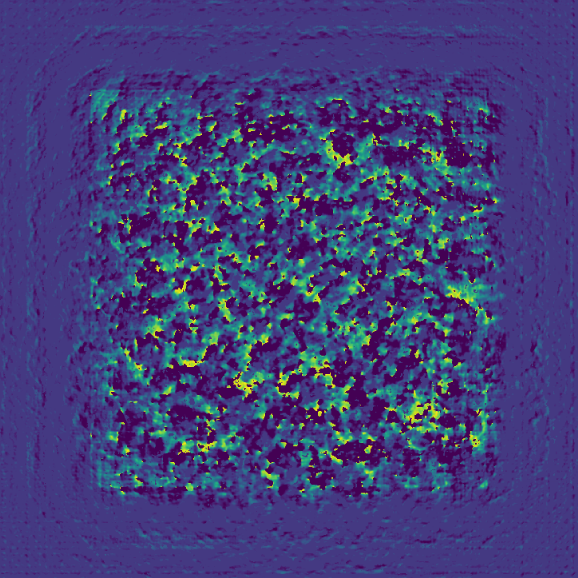}};
		\end{tikzpicture}       
		\begin{tikzpicture}
			\node[anchor=south west, inner sep=0] (image) at (0,0) {\includegraphics[width=0.24\columnwidth]{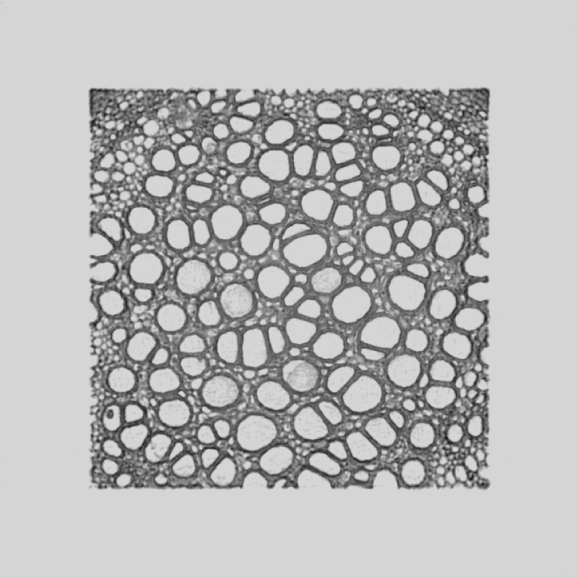}};
		\end{tikzpicture}      
		\begin{tikzpicture}
			\node[anchor=south west, inner sep=0] (image) at (0,0) {\includegraphics[width=0.24\columnwidth]{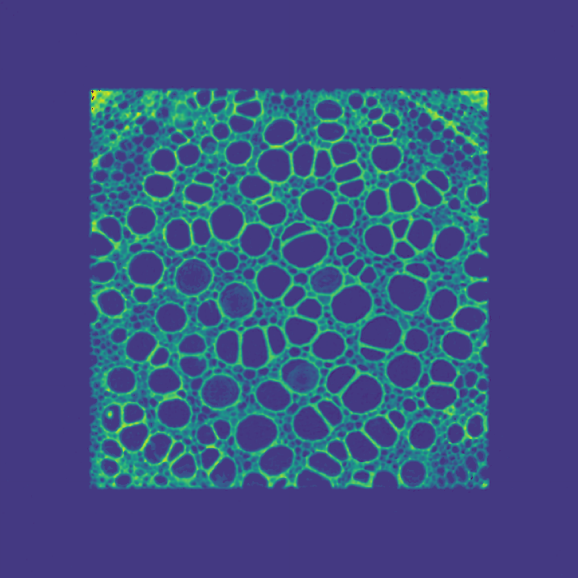}};
		\end{tikzpicture}  		
		\caption{Comparison results are presented for translation errors of 0 pixels, 11 pixels, and 22 pixels across the three planes.}
		\label{subfig:exp24}
	\end{subfigure}
	\caption{Comparison of reconstruction results using the ACC algorithm versus not using the ACC algorithm at different errors.}
	\label{fig:Experiment2}
\end{figure}
To further evaluate the ACC method's robustness against significant misalignment errors, we conducted a series of control experiments with varying degrees of misalignment. 
In these experiments, the first measurement plane acted as the reference, while the second measurement plane was incrementally displaced to the lower right by 2, 5, 8, and 11 pixels, and each subsequent measurement plane was shifted by double the amount of its predecessor. 
The comparisons between ACC-based reconstructions and those done using the naive approach are illustrated in Fig. \ref{fig:Experiment2}.
The results depicted in the figure indicate that as translational errors increase, the quality of reconstructions from the uncalibrated measurements significantly worsens. 
In contrast, reconstructions performed with ACC maintain consistent quality even as transnational errors become more pronounced, as reflected by their stable PSNR and SSIM scores in Table \ref{table2}. 
This demonstrates the proposed method's effectiveness in feature point detection and its capacity for robust calibration under scenarios of extensive misalignment.

\begin{table}[h]
	\caption{PSNR and SSIM of the reconstructions with various degree of translation errors as shown in Fig.~\ref{fig:Experiment2}.}\label{table2}
	\centering
    \begin{tabular}{ccccc}
	\hline
	  & \multicolumn{2}{c}{Na\"ive reconstruction} & \multicolumn{2}{c}{ACC reconstruction} \\
	\cline{2-5}
	Error/pixels & PSNR/dB & SSIM & PSNR/dB & SSIM\\
	\hline
	2 & 16.76 & 0.47 & 20.00 & 0.81 \\
	
	5 & 12.81 & 0.29 & 20.36 & 0.82 \\
	
	8 & 12.44 & 0.26 & 20.36 & 0.81 \\
	
	11 & 12.25 & 0.27 & 20.37 & 0.81 \\
	\hline
\end{tabular}
\end{table}

\subsection{Optical Experiments}

\begin{figure}[h]
	\centering
	\begin{subfigure}[t]{1\columnwidth}
		\centering
		\includegraphics[width=0.4\columnwidth]{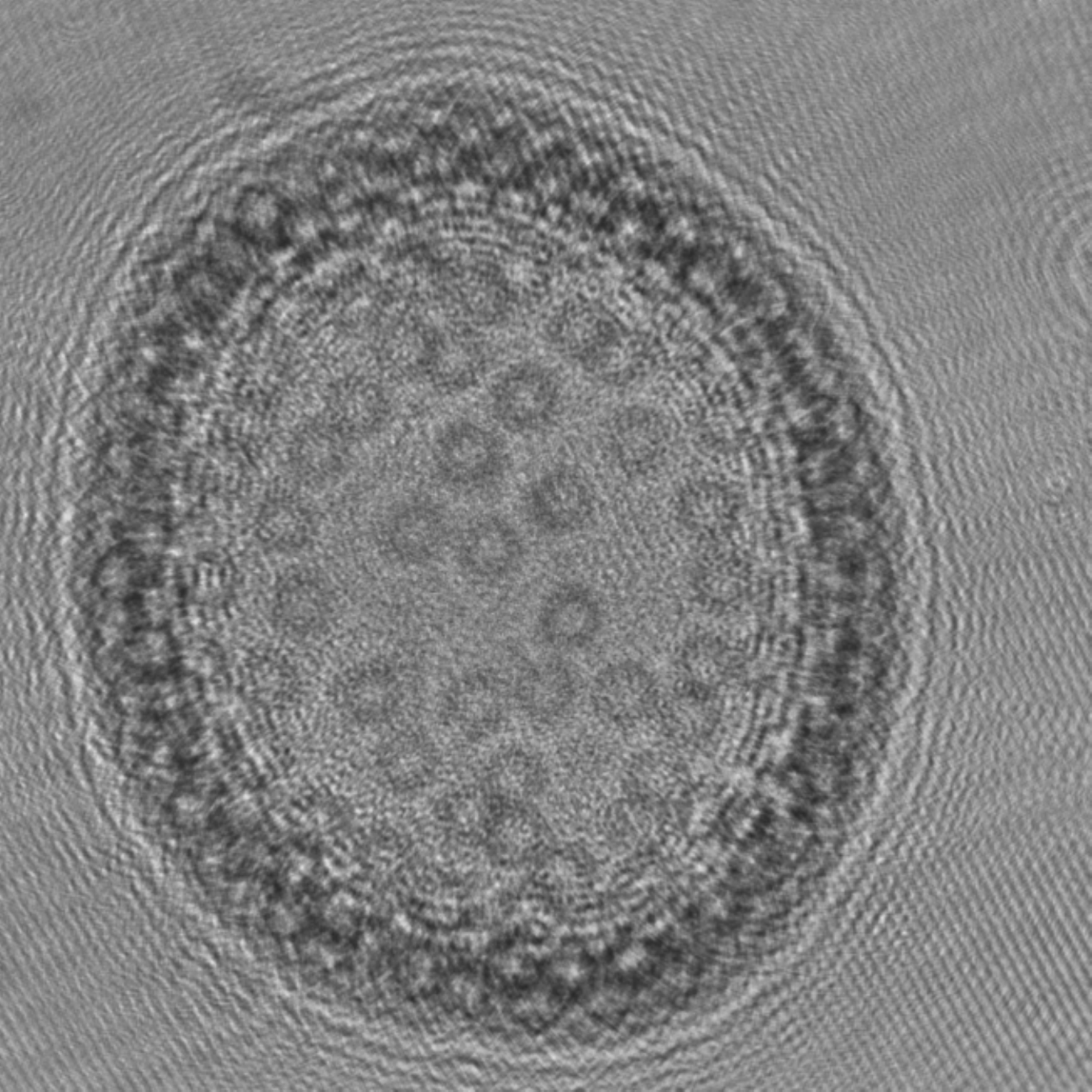}  
		\caption{Selected measurement}
		\label{subfig:plant_stems_holo}
	\end{subfigure}
 
	\begin{subfigure}[t]{1\columnwidth}
		\centering  
		\begin{tikzpicture}[spy using outlines={rectangle, width=0.7cm, height=0.7cm, spy_color_red, magnification=3, connect spies}]
			\node[anchor=south west, inner sep=0] (image) at (0,0) {\includegraphics[width=0.4\columnwidth]{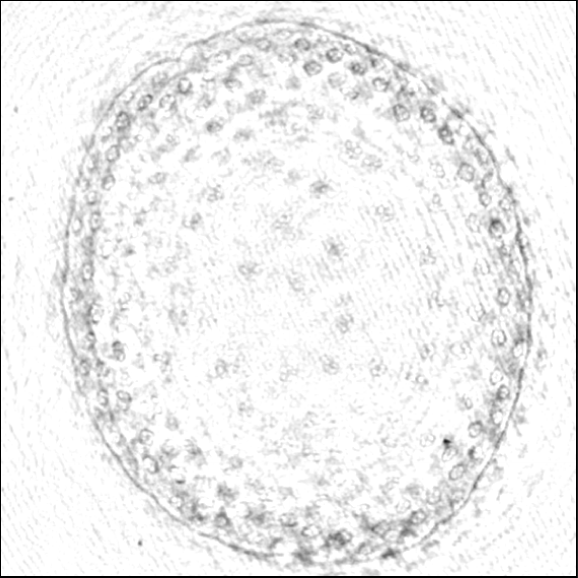}};
			\spy[spy_color_red, opacity=1, line width=0.5pt] on (2.95, 1.37) in node[anchor=south east, line width=0.5pt] at (3.32, 0.03);
			\spy[spy_color_red, opacity=1, line width=0.5pt] on (0.45, 2.05) in node[anchor=north west, line width=0.5pt] at  (0.03, 3.32);
			\node[align=left] at (1.5, 3.55) {amplitude};  
		\end{tikzpicture}    
		\begin{tikzpicture}[spy using outlines={rectangle, width=0.7cm, height=0.7cm, spy_color_red, magnification=3, connect spies}] 
			\node[anchor=south west, inner sep=0] (image) at (0,0) {\includegraphics[width=0.4\columnwidth]{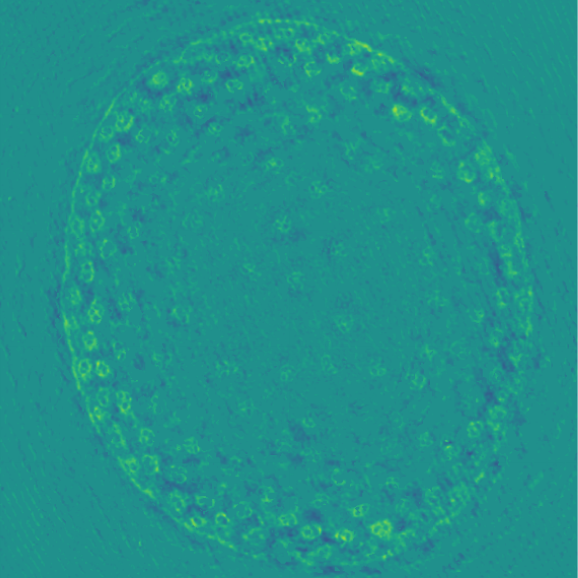}};
			\spy[spy_color_orange, opacity=1, line width=0.5pt] on (2.95, 1.37) in node[anchor=south east, line width=0.5pt] at (3.32, 0.03);
			\spy[spy_color_orange, opacity=1, line width=0.5pt] on (0.45, 2.05) in node[anchor=north west, line width=0.5pt] at  (0.03, 3.32);
			\node[align=left] at (1.5, 3.55) {phase};  
		\end{tikzpicture}     
		\caption{Nai\"ve reconstruction}
		\label{subfig:Uncalibrated plant stems}
	\end{subfigure} 
 
	\begin{subfigure}[t]{1\columnwidth}
		\centering
		\begin{tikzpicture}[spy using outlines={rectangle, width=0.7cm, height=0.7cm, spy_color_red, magnification=3, connect spies}]
			\node[anchor=south west, inner sep=0] (image) at (0,0) {\includegraphics[width=0.4\columnwidth]{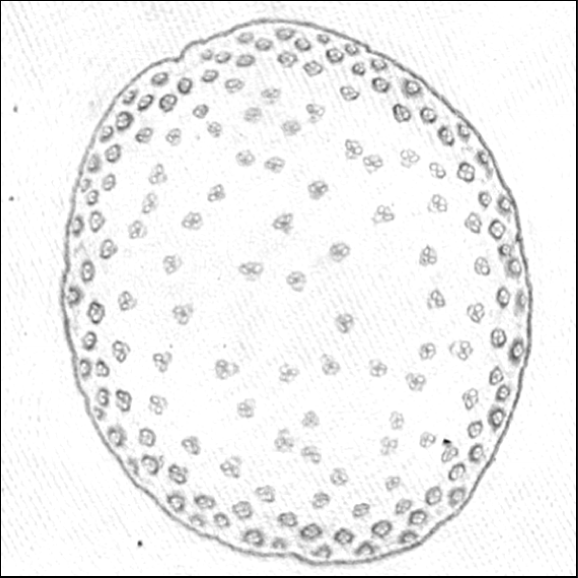}};
			\spy[spy_color_red, opacity=1, line width=0.5pt] on (2.95, 1.37) in node[anchor=south east, line width=0.5pt] at (3.32, 0.03);
			\spy[spy_color_red, opacity=1, line width=0.5pt] on (0.45, 2.05) in node[anchor=north west, line width=0.5pt] at  (0.03, 3.32);
			\node[align=left] at (1.5, 3.555) {amplitude};  
		\end{tikzpicture}
		\begin{tikzpicture}[spy using outlines={rectangle, width=0.7cm, height=0.7cm, spy_color_red, magnification=3, connect spies}]
			\node[anchor=south west, inner sep=0] (image) at (0,0) {\includegraphics[width=0.4\columnwidth]{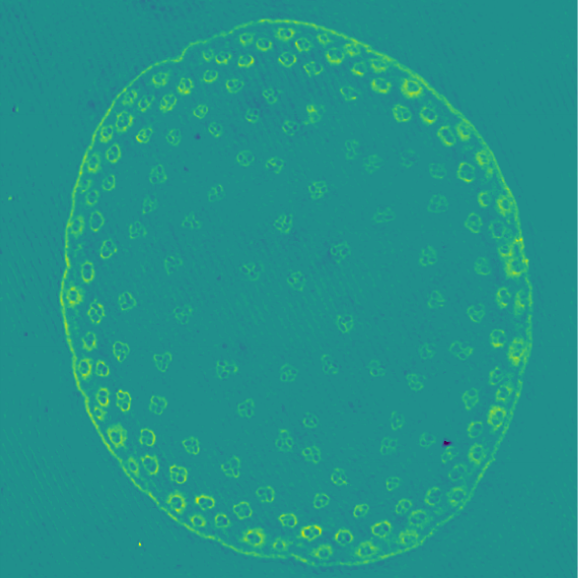}};
			\spy[spy_color_orange, opacity=1, line width=0.5pt] on (2.95, 1.37) in node[anchor=south east, line width=0.5pt] at (3.32, 0.03);
			\spy[spy_color_orange, opacity=1, line width=0.5pt] on (0.45, 2.05) in node[anchor=north west, line width=0.5pt] at (0.03, 3.32);  
			\node[align=left] at (1.5, 3.55) {phase}; 
		\end{tikzpicture}  
		\caption{ACC reconstruction}
		\label{subfig:Calibrated plant stems}
	\end{subfigure}
        \includegraphics[width=0.9\columnwidth]{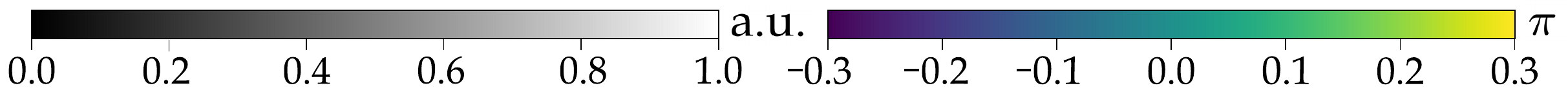}
	\caption{Reconstruction results of sample plant stems using the ACC algorithm versus not using the ACC algorithm.}
	\label{fig:Experiment3}
\end{figure}

We carry out experiments involving biological samples, following the schematic layout presented in Fig. \ref{error}. In these experiments, a laser emitting at a central wavelength of \SI{532}{\nano\meter} served as the illumination source. This laser light passed through an optical attenuator and a collimated beam expander before illuminating the biological specimens. An image sensor, with a pixel size of \SI{3.45}{\micro\meter}, captured the measurements. The distance between each measurement point was approximately \SI{5}{\milli\meter}.

\begin{figure}[h]
	\centering
        \begin{tikzpicture}
		\node[anchor=south west, inner sep=0] (image) at (0,0) {\includegraphics[width=3in]{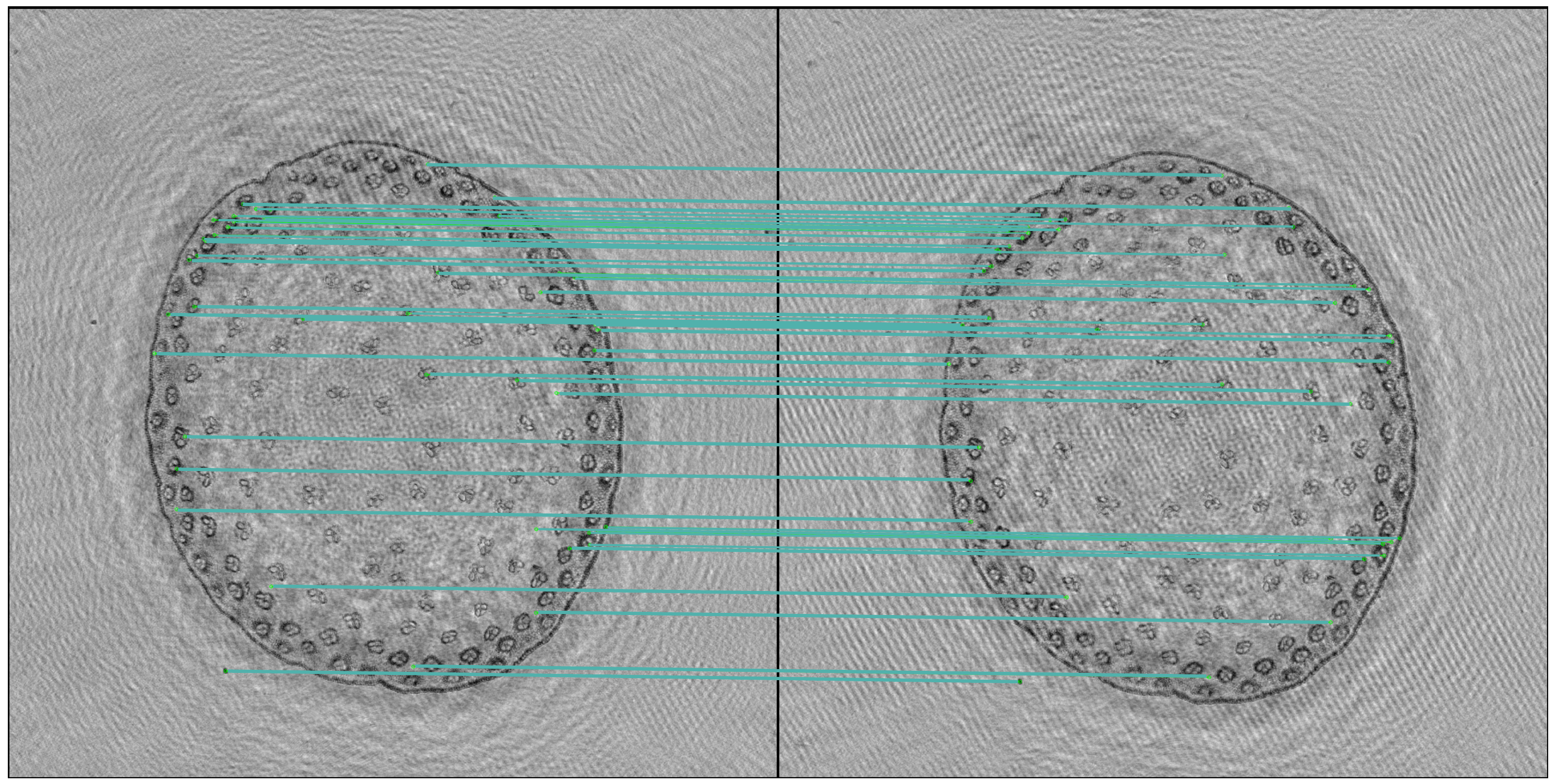}};
		\node[align=left, white] at (1.8, 3.6) {Measurement 2};  
            \node[align=left, white] at (5.8, 3.6) {Measurement 3};  
	\end{tikzpicture}
	\caption{Feature points in the refocused images of the plant stem sample.}
	\label{Detecting_exp}
\end{figure}
\begin{figure}[h]
	\centering
         \begin{tikzpicture}
		\node[anchor=south west, inner sep=0] (image) at (0,0) {\includegraphics[width=3in]{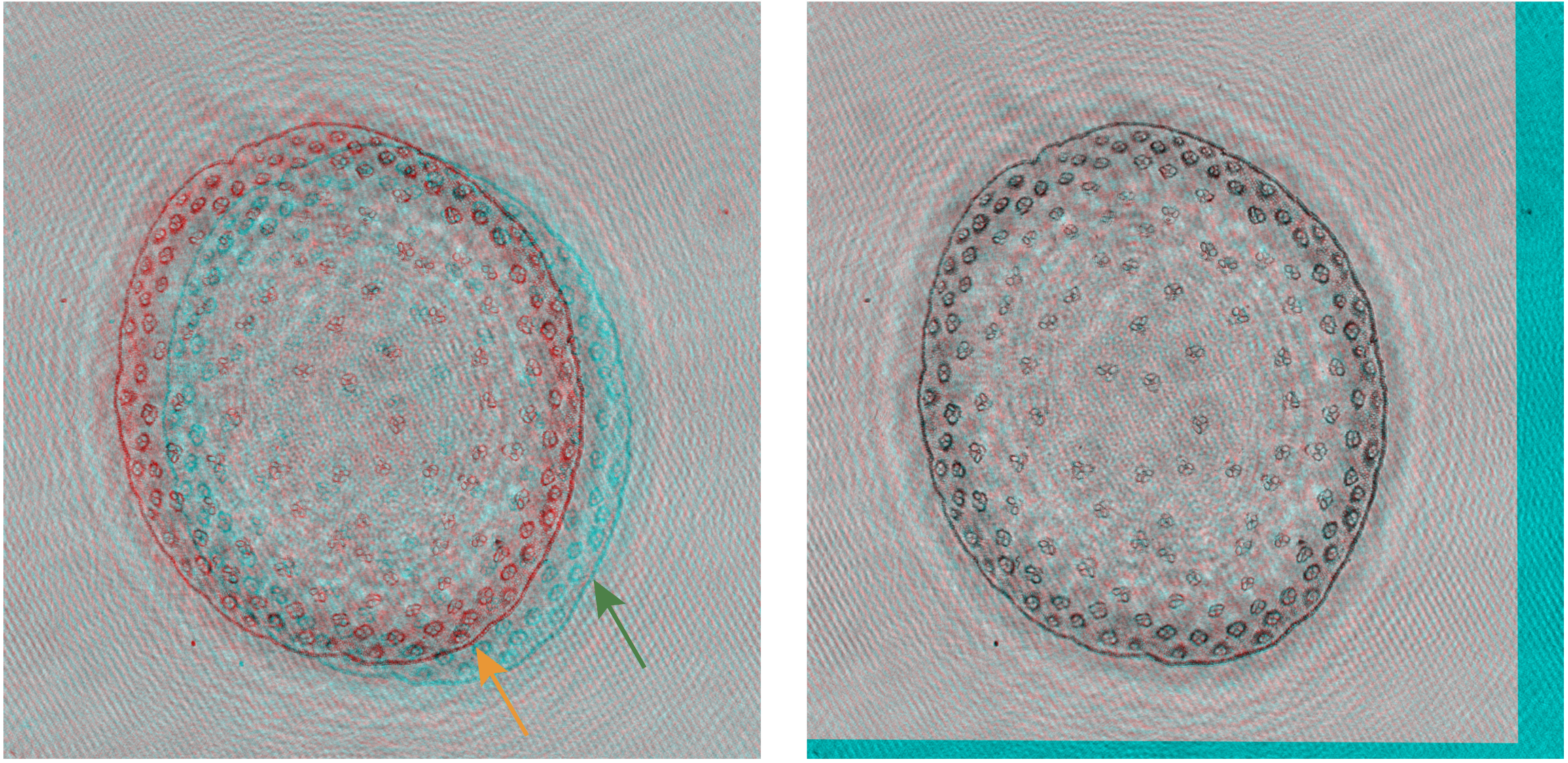}};
  		\node[align=left, white] at (1.8, 3.5) {Before calibration};
            \node[align=left, white] at (5.8, 3.5) {After calibration};
	\end{tikzpicture}
	\caption{Discrepancies between the first and third measurements before and after ACC calibration represented 
 by anaglyph.}
	\label{calibration_change}
\end{figure}

Figure \ref{fig:Experiment3}(a) presents a measurement from a plant stem sample, with the initial sample-to-sensor distance detected at \SI{34.1}{\milli\meter} by the auto-focusing algorithm. 
The results from the nai\"ve and ACC methods are depicted in Fig. \ref{fig:Experiment3}(b) and \ref{fig:Experiment3}(c), respectively. 
The nai\"ve reconstruction exhibits significant deterioration in comparison to the ACC reconstruction, which clearly delineates the cell structures.
Further analysis was conducted on the misalignment within the measurements. 
Figure \ref{Detecting_exp} illustrates the feature points identified and matched between the second and third measurements. 
To demonstrate the misalignment's extent before and after the calibration process, the first and third measurements are overlaid in different color channels by anaglyph in Fig. \ref{calibration_change}, accentuating the alignment discrepancies. 
In the left image, the first measurement indicated by a orange arrow and the third by a greed arrow.
The misalignment is visually evident in the left image before calibration, where ghosting effects from the overlay of two measurements indicate severe misalignment. 
Conversely, the right image in Fig. \ref{fig:Experiment3}(c) showcases the effective alignment correction achieved through the ACC method, highlighting its proficiency in addressing misalignment issues.

\begin{figure*}[h]
	\centering
	\begin{subfigure}[t]{0.3\columnwidth}
		\centering
		\includegraphics[width=1.16\columnwidth]{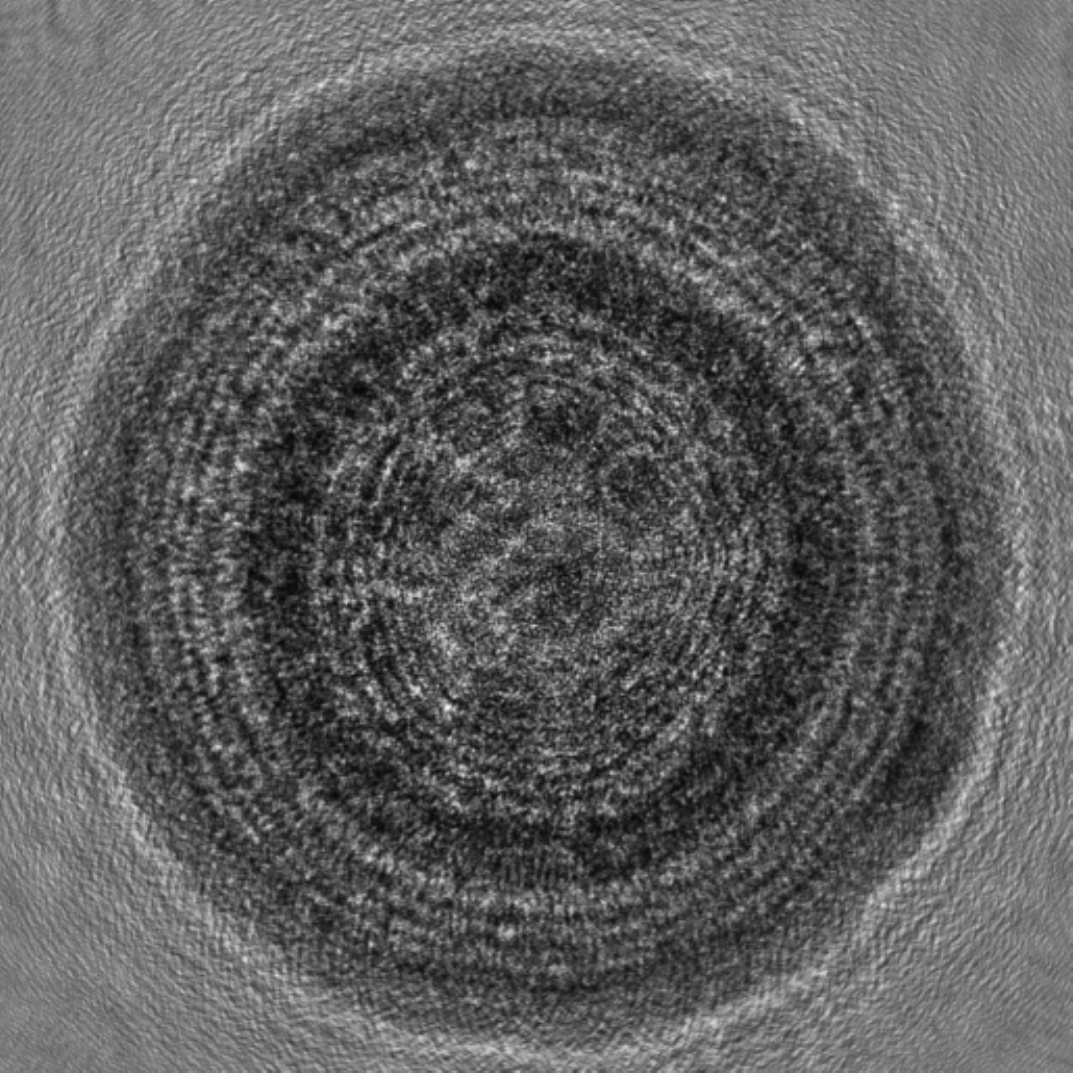}  
		\caption{measurement}
		\label{subfig:convallaria_holo}
	\end{subfigure}
        \hspace{0.00001\columnwidth} 
	\begin{subfigure}[t]{0.85\columnwidth}
		\centering  
		\begin{tikzpicture}[spy using outlines={rectangle, width=0.7cm, height=0.7cm, spy_color_red, magnification=3, connect spies}]
			\node[anchor=south west, inner sep=0] (image) at (0,0) {\includegraphics[width=0.41\columnwidth]{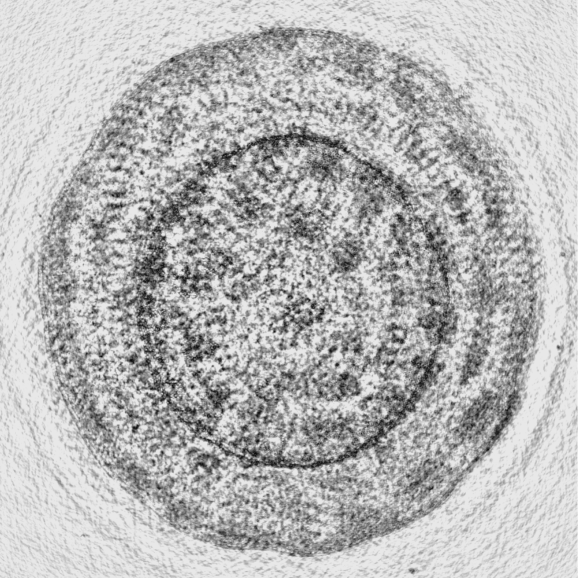}};
			\spy[spy_color_red, opacity=1, line width=0.5pt] on (1.8, 1.12) in node[anchor=south east, line width=0.5pt] at (2.9, 0.03);
			\spy[spy_color_red, opacity=1, line width=0.5pt] on (0.75, 1.55) in node[anchor=north west, line width=0.5pt] at (0.03, 2.9);
			\node[align=left] at (1.4, 3.2) {amplitude};   
		\end{tikzpicture}    
		\begin{tikzpicture}[spy using outlines={rectangle, width=0.7cm, height=0.7cm, spy_color_red, magnification=3, connect spies}] 
			\node[anchor=south west, inner sep=0] (image) at (0,0) {\includegraphics[width=0.41\columnwidth]{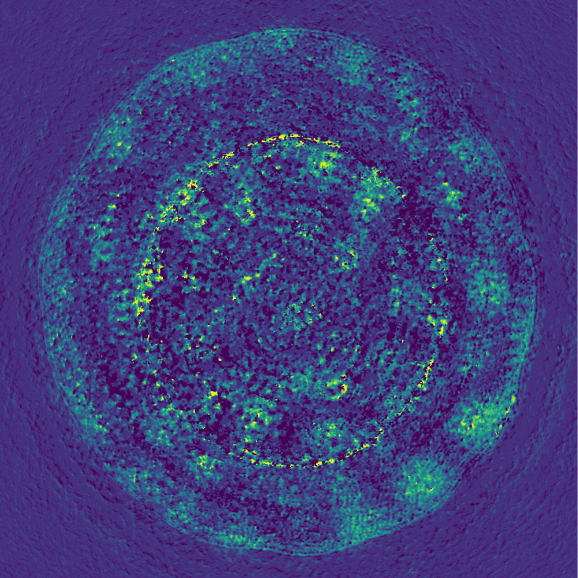}};
			\spy[spy_color_orange, opacity=1, line width=0.5pt] on (1.8, 1.12) in node[anchor=south east, line width=0.5pt] at (2.9, 0.03);
			\spy[spy_color_orange, opacity=1, line width=0.5pt] on (0.75, 1.55) in node[anchor=north west, line width=0.5pt] at (0.03, 2.9);
			\node[align=left] at (1.4, 3.2) {phase};  
		\end{tikzpicture}   
		\caption{Nai\"ve convallaria reconstruction}
		\label{subfig:Uncalibrated convallaria}
	\end{subfigure}
	\begin{subfigure}[t]{0.85\columnwidth}
		\centering
		\begin{tikzpicture}[spy using outlines={rectangle, width=0.7cm, height=0.7cm, spy_color_red, magnification=3, connect spies}]
			\node[anchor=south west, inner sep=0] (image) at (0,0) {\includegraphics[width=0.41\columnwidth]{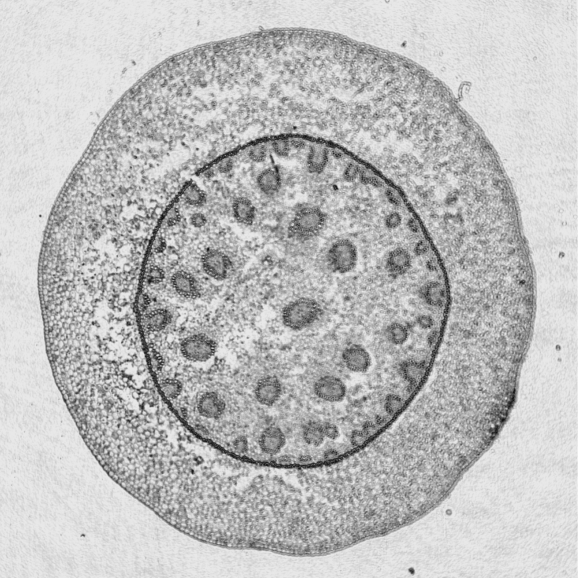}};
			\spy[spy_color_red, opacity=1, line width=0.5pt] on (1.8, 1.12) in node[anchor=south east, line width=0.5pt] at (2.9, 0.03);
			\spy[spy_color_red, opacity=1, line width=0.5pt] on (0.75, 1.55) in node[anchor=north west, line width=0.5pt] at (0.03, 2.9);
			\node[align=left] at (1.4, 3.2) {amplitude};  
		\end{tikzpicture}
		\begin{tikzpicture}[spy using outlines={rectangle, width=0.7cm, height=0.7cm, spy_color_red, magnification=3, connect spies}]
			\node[anchor=south west, inner sep=0] (image) at (0,0) {\includegraphics[width=0.41\columnwidth]{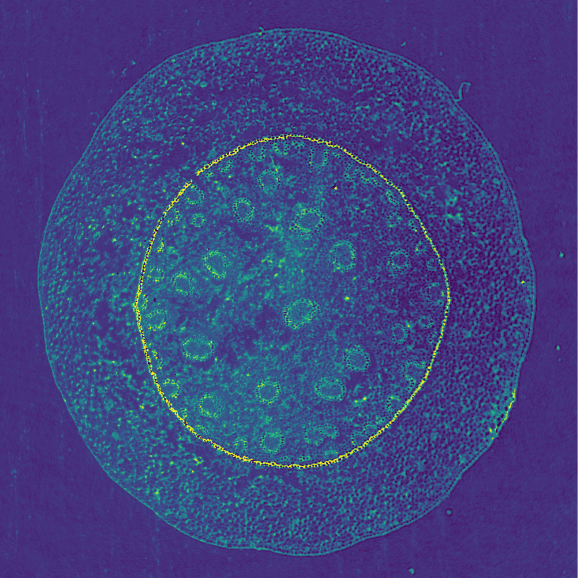}};
			\spy[spy_color_orange, opacity=1, line width=0.5pt] on (1.8, 1.12) in node[anchor=south east, line width=0.5pt] at (2.9, 0.03);
			\spy[spy_color_orange, opacity=1, line width=0.5pt] on (0.75, 1.55) in node[anchor=north west, line width=0.5pt] at (0.03, 2.9);   
			\node[align=left] at (1.4, 3.2) {phase};  
		\end{tikzpicture}
            \includegraphics[height=0.41\columnwidth]{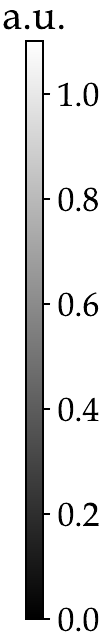} 
	    \includegraphics[height=0.41\columnwidth]{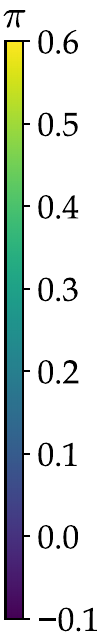}
		\caption{ACC convallaria reconstruction}
		\label{subfig:Calibrated convallaria}
	\end{subfigure}
 
	\begin{subfigure}[t]{0.3\columnwidth}
		\centering
		\includegraphics[width=1.16\columnwidth]{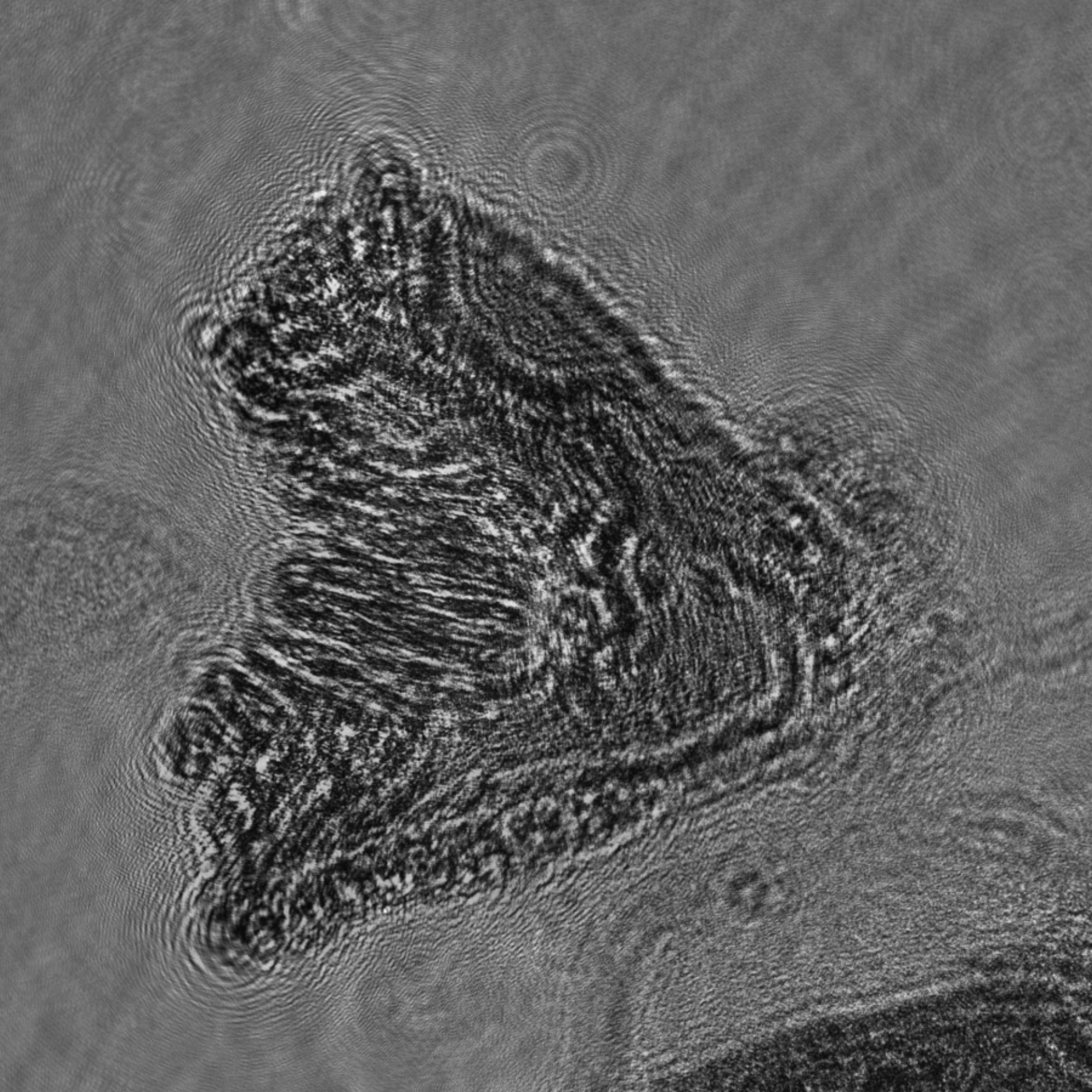}  
		\caption{measurement}
		\label{subfig:ruttus_holo}
	\end{subfigure}
	\begin{subfigure}[t]{0.85\columnwidth}
		\centering  
		\begin{tikzpicture}[spy using outlines={rectangle, width=0.7cm, height=0.7cm, spy_color_red, magnification=3, connect spies}]
			\node[anchor=south west, inner sep=0] (image) at (0,0) {\includegraphics[width=0.41\columnwidth]{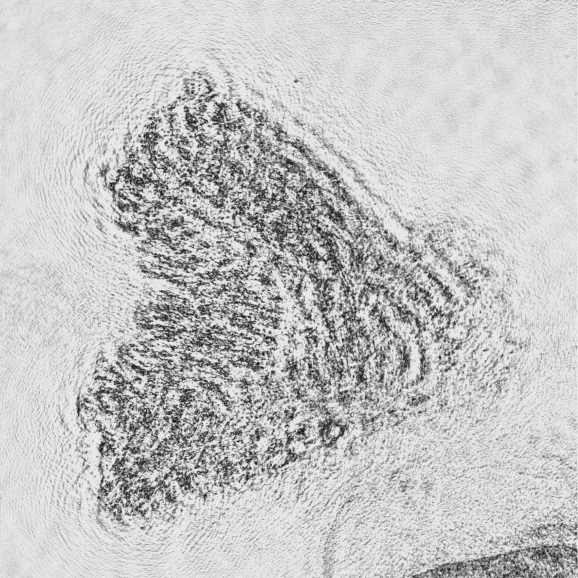}};
			\spy[spy_color_red, opacity=1, line width=0.5pt] on (1.7, 1.6) in node[anchor=south east, line width=0.5pt] at (2.9, 0.03);
			\spy[spy_color_red, opacity=1, line width=0.5pt] on (0.95, 2) in node[anchor=north west, line width=0.5pt] at (0.03, 2.9);
		\end{tikzpicture}    
		\begin{tikzpicture}[spy using outlines={rectangle, width=0.7cm, height=0.7cm, spy_color_red, magnification=3, connect spies}] 
			\node[anchor=south west, inner sep=0] (image) at (0,0) {\includegraphics[width=0.41\columnwidth]{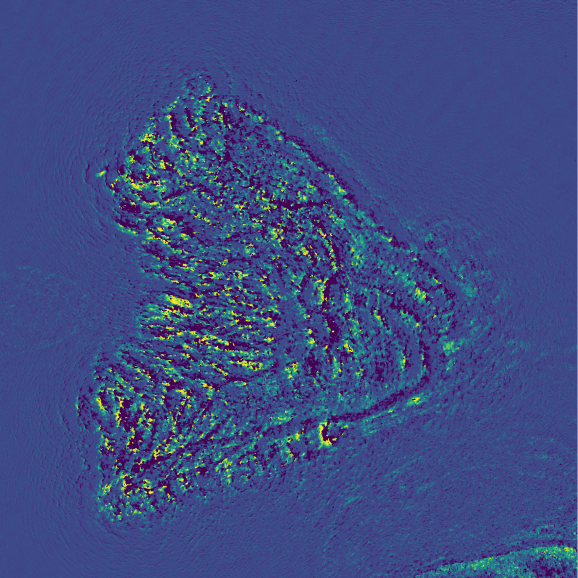}};
			\spy[spy_color_orange, opacity=1, line width=0.5pt] on (1.7, 1.6) in node[anchor=south east, line width=0.5pt] at (2.9, 0.03);
			\spy[spy_color_orange, opacity=1, line width=0.5pt] on (0.95, 2) in node[anchor=north west, line width=0.5pt] at (0.03, 2.9);
		\end{tikzpicture}     
		\caption{Nai\"ve ruttus reconstruction}
		\label{subfig:Uncalibrated ruttus}
	\end{subfigure} 
	\begin{subfigure}[t]{0.85\columnwidth}
		\centering
		\begin{tikzpicture}[spy using outlines={rectangle, width=0.7cm, height=0.7cm, spy_color_red, magnification=3, connect spies}]
			\node[anchor=south west, inner sep=0] (image) at (0,0) {\includegraphics[width=0.41\columnwidth]{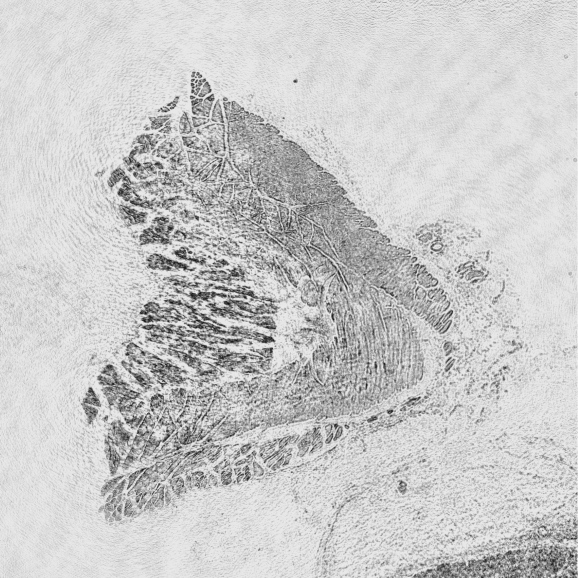}};
			\spy[spy_color_red, opacity=1, line width=0.5pt] on (1.7, 1.6) in node[anchor=south east, line width=0.5pt] at (2.9, 0.03);
			\spy[spy_color_red, opacity=1, line width=0.5pt] on (0.95, 2) in node[anchor=north west, line width=0.5pt] at (0.03, 2.9);
		\end{tikzpicture} 
		\begin{tikzpicture}[spy using outlines={rectangle, width=0.7cm, height=0.7cm, spy_color_red, magnification=3, connect spies}]
			\node[anchor=south west, inner sep=0] (image) at (0,0) {\includegraphics[width=0.41\columnwidth]{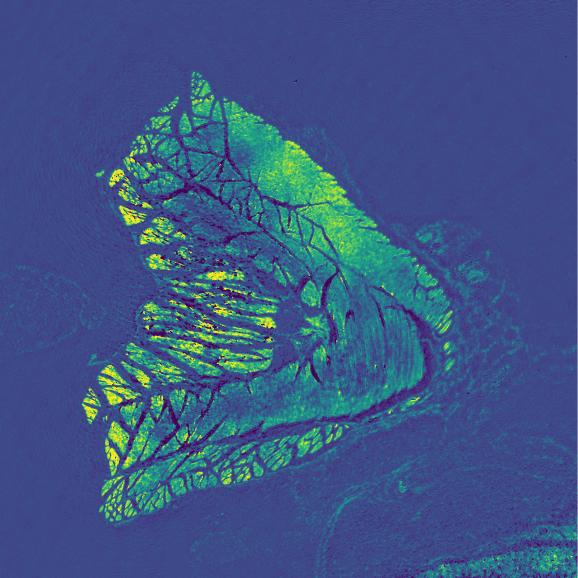}};
			\spy[spy_color_orange, opacity=1, line width=0.5pt] on (1.7, 1.6) in node[anchor=south east, line width=0.5pt] at (2.9, 0.03);
			\spy[spy_color_orange, opacity=1, line width=0.5pt] on (0.95, 2) in node[anchor=north west, line width=0.5pt] at (0.03, 2.9);   
		\end{tikzpicture}
            \includegraphics[height=0.41\columnwidth]{exp_mp_convallaria1_z0_amplitude_colorbar}  
	    \includegraphics[height=0.41\columnwidth]{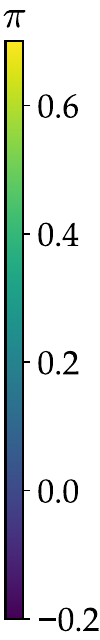}  
		\caption{ACC ruttus reconstruction}
		\label{subfig:Calibrated ruttus}
	\end{subfigure} 
	\caption{Comparison of reconstruction results for samples convallaria and ruttus using the ACC algorithm  versus not using the ACC algorithm.}
	\label{fig:Experiment4}
\end{figure*}

We conducted additional experiments focusing on two types of cells: convallaria, commonly known as lily of the valley, and rattus, or rat cells. The key measurements from these experiments are showcased in Fig. \ref{fig:Experiment4}, with figures (a) and (d) highlighting the initial captured images for convallaria and rattus cells, respectively.
In these experiments, we compare the reconstructions of using a na\"ve approach versus our ACC method. The results of the na\"ve reconstructions are depicted in Fig. \ref{fig:Experiment4}(b) for convallaria and (e) for rattus cells, illustrating the limitations of this approach in preserving the intricate details of the cellular structures.
Conversely, the ACC method's reconstructions, as shown in Fig.~\ref{fig:Experiment4}(c) for convallaria and (f) for rattus cells, demonstrate a significant improvement in clarity and detail. This method allows for a more accurate depiction of the cellular structures, enhancing our ability to discern the intricate details within the samples.
This pattern of results echoes the findings from previous experiments, such as those depicted in Fig. \ref{fig:Experiment3}, where the na\"ve reconstructions failed to accurately reveal the details of the samples. The ACC reconstructions, in contrast, successfully facilitate the visualization of clear and accurate cell structures. This further underscores the efficacy of the ACC method in overcoming the limitations of traditional reconstruction approaches, providing a more reliable means of analyzing cellular details in biological research.

\def\figwidth{0.1\columnwidth}

\section{Discussion}\label{sec:Discussion}
In the optical experiments depicted in Fig. \ref{fig:Experiment3}, the calibration matrices derived for measurement planes 2 and 3 of the plant stem samples are shown in Eq.~\ref{matrix_exp1} and Eq.~\ref{matrix_exp2}, respectively. Notably, the diagonal elements $a_{11}$, $a_{22}$, and $a_{33}$ in both matrices are close to $1$, indicating minimal scaling errors between these measurements. 
The values of $a_{12}$ and $a_{21}$, on the order of $10^{-3}$, suggest that rotational errors are also minor. Meanwhile, $a_{31}$ and $a_{32}$, with magnitudes around $10^{-6}$, are used to compute perspective transformations, showing their lesser significance. In contrast, the parameters $a_{13}$ and $a_{23}$, having an order of $10$, point out that translational errors are the most impactful.
This prominence of translational errors is attributed to the inline setup utilizing a plane wave for illumination. However, in different configurations, such as those involving spherical illumination or lensless setups, scaling factors cannot be overlooked.
\begin{equation}
	\label{matrix_exp1}
	\begin{bmatrix} 
		\quad9.957e^{-1}\quad  & -3.676e^{-3} & -4.416e^{1} \\ 
		\quad1.043e^{-3}\quad  & 9.965e^{-1}  & -2.053e^{1} \\
		\quad1.018e^{-7}\quad  & -4.421e^{-6} & 1
	\end{bmatrix}
\end{equation}

\begin{equation}
	\label{matrix_exp2}        
 	\begin{bmatrix} 
		9.894e^{-1}   &-3.094e^{-3}  & -4.239e^{1} \\ 
		-4.471e^{-3}  & 9.919e^{-1}  & -1.721e^{1} \\
		-6.079e^{-6}  & -2.786e^{-6} & 1
	\end{bmatrix}
\end{equation}

The proposed method outperforms traditional multi-plane phase retrieval approaches in accurately reconstructing measurements with larger errors.
However, it is structured in three sequential stages, where the outcome of each stage directly influences the overall reconstruction quality. For instance, an inability to accurately determine the reconstruction distance in the autofocusing algorithm may result in unclear reconstructions. 
Similarly, failure to precisely calibrate errors in the adaptive cascade calibration algorithm, or to correctly align multiple measurements, could severely degrade the quality of the final reconstructed image. Moreover, minor inaccuracies at each step can accumulate, further impacting the final result.
An integrated end-to-end process that can streamline these three steps into a single operation would greatly reduce the potential for cumulative errors, thereby enhancing the reliability and quality of the reconstruction outcome.

\section{Conclusion}
In conclusion, this study introduces the ACC method for multi-plane phase retrieval, specifically crafted to correct for misalignment encountered during the acquisition of measurements. This technique autonomously identifies feature points within the refocused sample space and iteratively computes the transformation matrix for adjacent planes, ensuring the calibration of all measurements. 
Experimental evidence substantiates the efficacy of this approach. 
Considering the common challenge of misalignment in computational imaging \cite{Chen2023}, it is expected that our strategy will provide a useful framework for enhancing various computational imaging techniques.

\verb+\printcredits+ command is used after appendix sections to list 
author credit taxonomy contribution roles tagged using \verb+\credit+ 
in frontmatter.

\printcredits


\bibliographystyle{model1-num-names}
\bibliography{reference}

\end{document}